\documentclass[
  reprint,
  superscriptaddress, 
  amsmath,amssymb,altaffilsymbol,
  aps,
  prx
]{revtex4-2}

\usepackage{graphicx}
\usepackage{dcolumn}
\usepackage{bm}

\usepackage{graphicx,epsf,bm,bbm,color,amssymb,float, subfigure}
\usepackage{amsmath,bm}
\usepackage{color}
\usepackage{textcomp}
\usepackage{dsfont}
\usepackage{tikz}
\usepackage{indentfirst}
\usepackage[colorlinks,citecolor=blue]{hyperref}
\usepackage{soul}
\usepackage{multirow}
\newcommand{\remove}[1]{}
\usepackage[most]{tcolorbox}

\newcommand{\supplementarysection}{%
  \setcounter{figure}{0}
  \let\oldthefigure\thefigure
  \renewcommand{\thefigure}{S\oldthefigure}
  \setcounter{section}{0}
  \let\oldthesection\thesection
  \renewcommand{\thesection}{S\oldthesection}
  \setcounter{equation}{0}
  \let\oldtheequation\theequation
  \renewcommand{\theequation}{S\oldtheequation}
  \setcounter{table}{0}
  \let\oldthetable\thetable
  \renewcommand{\thetable}{S\oldthetable}
}

\newcommand{\bi}{\begin{itemize}}
\newcommand{\ei}{\end{itemize}}
\newcommand{\be}{\begin{enumerate}}
\newcommand{\ee}{\end{enumerate}}

\newenvironment{dfn}{{\vspace*{1ex} \noindent \bf Definition }}{\vspace*{1ex}}

\newcommand{\ie}{{\it i.e.}}
\newcommand{\Eq}[1]{Eq.~(\ref{#1})}

\newcommand{\trace}[1]{\mathrm{tr}{\left(#1\right)}}
	
	\newcommand{\beq}{\begin{eqnarray}}
	\newcommand{\eeq}{\end{eqnarray}}
	\newcommand{\bea}{\begin{eqnarray}\begin{aligned}}
	\newcommand{\eea}{\end{aligned}\end{eqnarray}}

\usepackage[most]{tcolorbox}

\newtcolorbox{resultbox}{
  colback=white,
  colframe=black,
  boxrule=0.6pt,
  arc=0pt,
  left=4pt,
  right=4pt,
  top=4pt,
  bottom=4pt
}

\makeatletter
\def\@fnsymbol#1{%
  \ensuremath{%
    \ifcase#1
    \or *                
    \or \sharp           
    \else\@ctrerr
    \fi}}
\makeatother

\definecolor{forestgreen}{RGB}{34,139,34}

\begin{document}

\title{SC$^\sharp$: superconductivity intertwined with topological order}

\author{Zhi-Qiang Gao}
\affiliation{Department of Physics, University of California, Berkeley, California 94720, USA}
\affiliation{Department of Physics, Harvard University, Cambridge, Massachusetts 02138, USA}
\author{Yan-Qi Wang}
\affiliation{Department of Physics and Joint Quantum Institute, University of Maryland, College Park, Maryland 20742, USA}
\author{Hui Yang}
\affiliation{Department of Physics and Astronomy, University of Pittsburgh, Pennsylvania 15213, USA}
\author{Zhaoyu Han}
\email{zhan@fas.harvard.edu}
\affiliation{Department of Physics, Harvard University, Cambridge, Massachusetts 02138, USA}
\author{Ashvin Vishwanath}
\email{avishwanath@g.harvard.edu}
\affiliation{Department of Physics, Harvard University, Cambridge, Massachusetts 02138, USA}

\begin{abstract}

Superconductivity and intrinsic topological order can intertwine in ways that go beyond simple coexistence. Here we develop a classification framework for (2+1)D gapped superconductors with condensate charge $2ne$. By focusing on the topological properties of superconducting vortices, we precisely characterize this interplay. We obtain a classification with increasing levels of intertwinement, from the well known invertible superconductors without topological order, to untwisted and twisted SC$^\ast$, and finally to SC$^{\sharp}$. Our main finding is the family SC$^{\sharp}$, in which the elementary vortex carrying $h/(2ne)$ flux cannot be factorized into an invertible superconductor vortex dressed by bulk anyons, and every elementary vortex is necessarily non-Abelian. For such phases, we also obtain no-go results: the elementary vortex sector has quantum dimension at least $\sqrt{6}$ and cannot satisfy Fibonacci fusion rules. We develop a systematic algebraic framework to obtain all these superconducting phases through anyon condensation, allowing us to construct several infinite families of examples and make connections to anyon superconductivity. Finally, we introduce primitive SC$^\sharp$ phases for which factorization of every nontrivial vortex sector is obstructed, and argue that they cannot arise by disordering the relative phases of lower-charge superconducting condensates. Our classification scheme based on vortex properties allows us to uncover new distinctions between topologically ordered superconductors and constrain how they can be realized.

\end{abstract}

\maketitle

\tableofcontents

\section{Introduction}

Superconductivity (SC) and topological order (TO) are two paradigmatic macroscopic quantum phenomena that have long played central roles in condensed matter physics and quantum technology. It is known that the properties of SC, especially those of its vortices, can be qualitatively modified by topology in (2+1)D. A canonical example is the charge-\(2e\) \((p\!+\!ip)\)-wave weak-pairing SC in two spatial dimensions, whose elementary vortex traps a Majorana zero mode. More formally, this state, together with other charge-\(2e\) SC effectively described by noninteracting Bogoliubov--de Gennes (BdG) Hamiltonians, provides an example of a fermionic \(\mathbb{Z}_2\) symmetry-protected topological (SPT) phase, or invertible fermionic topological phase~\cite{Gu2014}. Such phases are classified by Kitaev's 16-fold way~\cite{16foldway} and have long been considered promising platforms for fault-tolerant quantum computation~\cite{Nayak2008}. Beyond these simplest examples, SC can interplay with topology in much richer ways. For example, there are SC phases that can be viewed as an invertible SC stacked with an intrinsic TO, often referred to as \(\mathrm{SC}^*\)~\cite{Senthil2000,Senthil2001,PhysRevLett.111.157004,PhysRevB.88.195117,Bunney2025,tang2025,Kim2025}.

More generally, SC need not originate from the condensation of ordinary charge-\(2e\) Cooper pairs. The condensed object may instead consist of more than two electrons, giving rise to a charge-\(2ne\) SC when the order parameter carries electric charge \(2ne\). For \(n>1\), such an order parameter indicates intrinsically correlated SC beyond a free-fermion BdG description. Nevertheless, a complete classification of  invertible charge-$2ne$ SC as fermionic  $\mathbb{Z}_{2n}$-SPT has been obtained~\cite{Wang20102me}. Here we will be interested in extending this discussion beyond invertible phases, allowing for the possibility of intrinsic TO accompanying the charge-$2ne$ SC. The aforementioned simple coexistence between TO and SC, namely SC$^\ast$, is only the simplest possibility. As we classify and describe below, TO and SC can intertwine in much deeper ways, as exemplified in recent studies~\cite{Gao2026,shi2025nonAbelian,Shi2026,Wang20266e}. 

We note several physical mechanisms can give rise to higher-charge SCs. One important route is through vestigial order of multi-component charge-$2e$ SC orders. That is, fluctuations may destroy the charge-$2e$ condensate while preserving a composite higher-charge condensate. This mechanism has been extensively discussed for pair density wave states~\cite{Berg2009,Radzihovsky2009,Radzihovsky2011} and other multicomponent SC~\cite{Vidal2002,Babaev2004,Radzihovsky2004,Radzihovsky2008,Herland2010,Zhou2021,Jian2021,Fernandes2021,Grinenko2021,Liu2023,Babaev2024,Zeng2024,Volovik2024,Soldini2024,Dai2024,samoilenka2025,zou2025,Shi2026,Pan2026}. It was also proposed that the nontrivial center of certain Lie group symmetries can play an essential role in stabilizing higher charge SC at zero temperature~\cite{Gao20264e,Wang20266e}, generalizing the long-standing proposal of quantum charge-$4e$ SC in $SU(4)$-symmetric systems~\cite{Wu2005,Khalaf2022,Lecheminant2005,Capponi2007,Capponi2008,Roux2008,Ko2008,Zhang2020,Wu2024,Zhang20254e,Li20264e,Wan2026}. 

Recently, several exotic examples of higher-charge SC with coexisting TO have been constructed, in which SC vortices bind non-Abelian zero modes that are absent from the defect-free sector~\cite{Gao2026,Shi2026}. These zero modes are more intricate than Majorana zero modes, indicating that such systems cannot be understood as simple stacks of an invertible SC and a TO. Instead, they reveal a fundamentally new possibility: SC and intrinsic TO can be {\it topologically intertwined} in an irreducible way.

This observation motivates a natural question: how can one systematically classify SC phases according to their topological properties and their possible intertwining with the symmetry-breaking order? To disambiguate, by SC we mean a spontaneous symmetry breaking order reducing the charge $U(1)$ symmetry down to a residual subgroup $\mathbb{Z}_{2n}$ containing the fermion parity, whose elementary defects are the vortices of the condensation field. Those defects are logarithmically confined below the BKT transition temperature, and can be induced by external $h/(2ne)$ fluxes. The full topological data of the system include both deconfined bulk anyons and the vortices. For classification purposes, we gap the  Goldstone modes with a weak symmetry breaking field and consider the resulting gapped ground state.  

One approach is to treat SC as a fermionic symmetry-enriched topological (SET) phase with $\mathbb{Z}_{2n}$ symmetry and apply the algebraic machinery of Refs.~\cite{BBCW,Bulmash2022} to classify the resulting fermionic SETs.  In this work, we instead adopt a physical perspective tailored to SC. We propose an organizing principle for classifying all (2+1)-dimensional SC phases with gapped $1,\dots,(2n\!-\!1)$-body spectra based on the topological properties of vortices. Within this framework, we characterize the degree of intertwinement between SC and TO across several distinct levels, as shown in Fig.~\ref{fig:hierarchy}. Among these classes, we identify a particularly interesting possibility where the elementary vortex sector cannot be factorized into a product of the deconfined sector and an elementary vortex sector of an invertible SC. Anyon permutation by the residual \(\mathbb{Z}_{2n}\) symmetry is sufficient, but {\em not necessary} for this nonfactorization. We call these phases SC$^\sharp$ (pronounced ``SC-sharp'') to suggest a heightened intertwinement between SC and TO, and to distinguish them from conventional SC$^\ast$ formed by stacking SC on an intrinsic TO. This definition has immediate consequences; for example, the elementary vortex sector of an SC$^\sharp$ has quantum dimension at least \(\sqrt{6}\), and cannot contain certain kinds of vortices, in particular the vortices with Abelian or Fibonacci quantum dimension. Besides the examples discovered in Refs.~\cite{Gao2026,Shi2026}, we construct several infinite families of new examples of such states.

We further show that our classification is exhaustive by proving that every (2+1)-dimensional SC phase can be obtained via condensation of a single self-bosonic anyon~\cite{Bais2009,KONG2014,hu2022anyon,AnyonCondensationReview} from a parent TO. This perspective is closely related in spirit to anyon superconductivity~\cite{FetterHannaLaughlin,Laughlin1988,Lee1989,WilczekWittenHalperinAnyonSC,Senthil2025,Divic2025,Pichler2025,shi2025nonAbelian,Seo2026} in the stack and condense picture \cite{Zhang2025,shi2025nonAbelian}. We note that such kinematic mappings can be achieved dynamically as anyon proliferation transitions~\cite{ChengSeiberg,cheng2026peelingAbeliananyonsproliferation}. 

Finally, we argue that the topological content of a charge-$2ne$ SC constrains its possible microscopic mechanism. In particular, the smallest $r\mid 2n$ rendering the \(rh/(2ne)\) vortex sector factorizable indicates that the coexisting TO is enriched only by the \(\mathbb{Z}_{r}\) subgroup of the residual \(\mathbb{Z}_{2n}\) symmetry. This suggests that the SC state could be viewed as formed by topologically binding charge-$ke$ ($k\equiv2n/r$) topologically trivial bound clusters of electrons. Consequently, such a state cannot arise  by partially melting a multicomponent SC with condensate charge greater than $ke$, {\it i.e.}, from a vestigial-order mechanism. In the extreme case \(k=1\), the SC$^\sharp$ phase cannot arise as a quantum vestigial order at all, and hence could only be a ``primary order." We hence call those states primitive SC$^\sharp$.

The paper is organized as follows. In Sec.~\ref{sec:sum} we summarize the main results of the paper. In Sec.~\ref{sec:review} we review the algebraic theory of anyons and symmetry defects. In Sec.~\ref{sec:classification} we introduce the classification framework of SC based on its vortex sector. In Sec.~\ref{sec:construct} we present the unified construction of SC based on anyon condensation. In Sec.~\ref{sec:mechanism} we discuss the relation of our classification and construction to the physical mechanisms of higher-charge SC, including vestigial order and anyon superconductivity. In Sec.~\ref{sec:ex} we construct examples using anyon condensation, including several (primitive) series of SC$^\sharp$. Sec.~\ref{sec:outlook} is devoted to conclusions and perspectives.

\section{Overview}\label{sec:sum}

Throughout this work, we use the following notation. The category containing all deconfined excitations, including the physical electron as the transparent fermion, is called the \emph{deconfined sector} and is denoted by $\mathcal{C}_{\mathfrak{0}}$. A useful notation is a reduced anyon set by modding out the physical electron $\mathcal{C}^\mathrm{dym}=\mathcal{C}_\mathfrak{0}/\mathrm{sVec}$, which will be called the \emph{dynamical anyon sector}, as the physical electron is typically not viewed as an anyon. Here $\mathrm{sVec}=\{1,f\}$ consists of the vacuum $1$ and the physical electron $f$. The dynamical anyon sector $\mathcal{C}^\mathrm{dym}$ need not be a category, as the fusion of objects in $\mathcal{C}^\mathrm{dym}$ may not be closed. We denote the $ph/(2ne)$ vortex sector of a charge-$2ne$ SC, composed of all topologically inequivalent SC vortices carrying flux $ph/(2ne)$, by $\mathcal{C}_{\mathfrak{p}}$ with $p\in\mathbb{Z}_{2n}$. In particular, the elementary vortex sector is $\mathcal{C}_{\mathfrak{1}}$, which is of central importance to the classification. We also introduce $\mathcal{C}_{\mathfrak{p}}^{\mathrm{inv}}$ for the $ph/(2ne)$ vortex sector of any one class of the charge-$2ne$ invertible SC~\cite{16foldway,Wang20102me}. A sector formed by every vortex in $\mathcal{C}_{\mathfrak{p}}^{\mathrm{inv}}$ fused with a topologically nontrivial Abelian object $v$ is denoted as $\mathcal{V}_\mathfrak{p}^\mathrm{inv}$.

\begin{figure}[t]
  \centering
  \includegraphics[width=\linewidth]{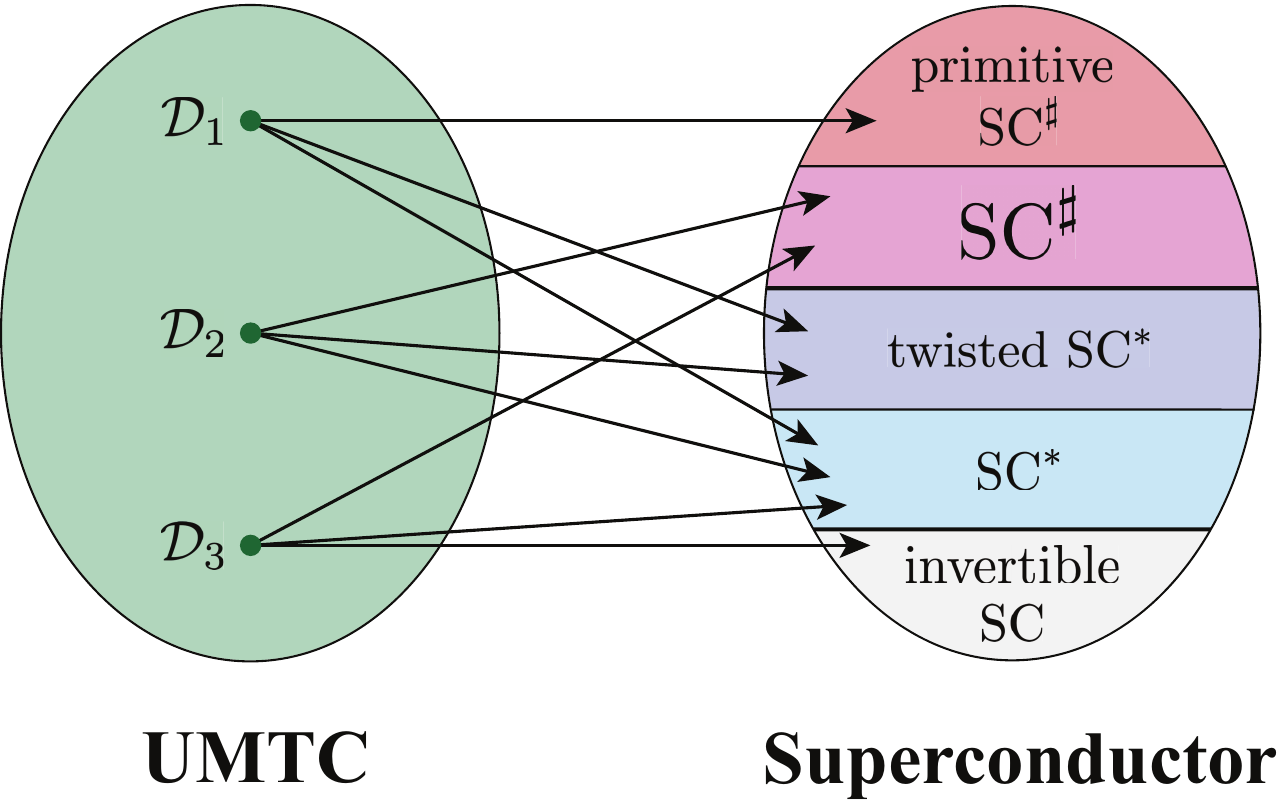}
  \caption{{\bf Grading the intertwinement between superconductivity (SC) and topological order (TO).} Distinct classes are organized according to the topological properties of vortices. All five classes can be constructed from anyon condensation in some parent UMTC $\mathcal{D}$ stacked with an atomic insulator, with a many-to-many correspondence in general.}
  \label{fig:hierarchy}
\end{figure}

{\em Classification based on vortex sectors}: Charge-$2ne$ SC with gapped $1,\dots,(2n\!-\!1)$-body spectra are organized into several broad classes according to their vortex sectors, especially the elementary vortex sector, as follows. See Fig.~\ref{fig:hierarchy} for illustration.

\begin{itemize}
    \item Invertible SC. A superconductor is called invertible if its elementary vortex sector is $\mathcal{C}_\mathfrak{1}^\mathrm{inv}$, meaning that it is an invertible phase with trivial $\mathcal{C}^\mathrm{dym}$. See Sec.~\ref{subsec:SC1}.
    \item (Untwisted) SC$^\ast$. A superconductor is called SC$^\ast$ if its elementary vortex sector is factorized as $\mathcal{C}_\mathfrak{1}^\mathrm{inv}\otimes\mathcal{C}^\mathrm{dym}$, suggesting that it is a stacking of an invertible SC and a neutral TO. See Sec.~\ref{subsec:SC2}.
    \item Twisted SC$^\ast$. A superconductor is called twisted SC$^\ast$ if its elementary vortex sector is factorized as $\mathcal{V}_\mathfrak{1}^\mathrm{inv}\otimes\mathcal{C}^\mathrm{dym}$. Here the Abelian object $v$ satisfies $v^{2n}=w\neq 1$ with $w\in\mathcal{C}^\mathrm{dym}$, while $v$ itself does not necessarily belong to $\mathcal{C}^\mathrm{dym}$. See Sec.~\ref{subsec:SC3}.
    \item SC$^\sharp$. A superconductor is called SC$^\sharp$ if its elementary vortex sector cannot be factorized as $\mathcal{C}_\mathfrak{1}^\mathrm{inv}\otimes\mathcal{C}^\mathrm{dym}$ or $\mathcal{V}_\mathfrak{1}^\mathrm{inv}\otimes\mathcal{C}^\mathrm{dym}$. SC$^\sharp$ defines a new class of SC in which the superconducting order and the topological order are irreducibly intertwined, and is of central interest to this work. See Sec.~\ref{subsec:SC4}.
    \item Primitive SC$^\sharp$. An SC$^\sharp$ is called primitive if all of its $ph/(2ne)$ vortex sectors cannot be factorized as $\mathcal{C}_\mathfrak{p}^\mathrm{inv}\otimes\mathcal{C}^\mathrm{dym}$ or $\mathcal{V}_\mathfrak{p}^\mathrm{inv}\otimes\mathcal{C}^\mathrm{dym}$. See Sec.~\ref{subsec:SC5}.
\end{itemize}

{\em General Construction}: Every SC can be constructed via anyon condensation from a parent UMTC $\mathcal{D}$ stacked with an atomic insulator $\mathrm{sRep}(U(1))$. By {\em atomic insulator}, we simply mean a trivial electronic insulator without anyons, that can be continuously deformed to a product state and serves as a reservoir of integer-charged excitations with charge ${k}e$ where $k$ even (odd) corresponds to bosonic (fermionic) excitations. The recipe is summarized as follows. First, search for an Abelian anyon $g\in\mathcal{D}$, with with either bosonic or fermionic self-statistics and denote the order of $g$ as $r$, \ie, the smallest positive integer yielding $g^r=1$. Second, stack $\mathcal{D}$ with an atomic insulator with local objects $\mathbf{k}$ labeled by their electric charge $ke$. 
Third, condense the bosonic composite $(g,\mathbf{k})$, which automatically gives rise to a charge-$2ne$ SC with $2n=kr$. The deconfined anyons in this SC correspond to those anyons in $\mathcal{D}$ having trivial monodromy with $g$, and the $ph/(2ne)$ vortices to those with monodromy $M=e^{2\pi i\frac{kp}{2n}}$ for $p=1,\cdots,r\!-\!1$. Both anyons and vortices are subject to identification by $g$ and possible splitting during the condensation of $(g,\mathbf{k})$.  Vortices carrying flux quanta in multiples of $h/(ke)$ are trivial, in the sense that they are formed by fusing deconfined excitations with a bare Abelian vortex with trivial topological spin. In particular, when the $h/(2e)$ vortices are trivial, such a SC can be viewed as a bosonic superfluid stacked with a trivial charge-$2ne$ SC. In this construction, all the topological data of the charge-$2ne$ SC are derivable from the parent UMTC and the condensed anyon $g$. See Sec.~\ref{sec:construct}.

\section{Review of the algebraic theory of anyons and symmetry defects}\label{sec:review}

The universal properties of a gapped {\it bosonic} TO---its anyon types, fusion rules, braiding statistics, and chiral central charge---are encoded by a unitary modular tensor category (UMTC). This framework unifies many physical descriptions, including Chern--Simons (CS) theories, Dijkgraaf--Witten theory~\cite{DWT}, exactly solvable lattice models~\cite{QD,16foldway}, and edge conformal field theories, and provides a precise way to distinguish phases that cannot be characterized by local order parameters alone. Its simple objects, denoted by $a,b,c,\cdots$, represent anyon types and obey fusion rules $a\times b=\sum_c N_{ab}^c \,c$, with non-negative integer coefficients $N_{ab}^c$. Each simple object has a quantum dimension $d_a$ and topological spin $s_a\in [0,1)$. The quantum dimension of anyons is determined by the fusion rule, $d_ad_b=\sum_c N_{ab}^c \, d_c$, and the total quantum dimension of $\mathcal{D}$ is $d_\mathcal{D}^2=\sum_{a\in\mathcal{D}}d_a^2$. The full fusion and braiding structure is specified by the fusion rules together with $F$- and $R$-symbols satisfying the pentagon and hexagon equations. Gauge invariant information is encoded in the modular matrices $S$ and $T$. Together with the chiral central charge modulo $8$, they determine the full modular data of a UMTC~\cite{Ng2023}.  

The structure becomes richer with global symmetry. A symmetry can permute anyons, endow them with fractional quantum numbers, and give rise to symmetry defects whose fusion and braiding extend the original UMTC data. The resulting framework of bosonic symmetry-enriched topological (bSET) phases describes how TO and symmetry intertwine~\cite{BBCW}. In the special case where the underlying UMTC is trivial, this framework reduces to the theory of SPT phases. More generally, for a TO described by a UMTC $\mathcal{D}$, enrichment by a symmetry group $G$ is described by a $G$-graded fusion category $\mathcal{D}_G=\bigoplus_{\mathfrak{g}\in G} \mathcal{D}_\mathfrak{g}$~\cite{BBCW}. The identity component $\mathcal{D}_\mathfrak{0}=\mathcal{D}$ describes the deconfined anyons, and $\mathcal{D}_\mathfrak{g}$ is associated with the defect sector of the group element $\mathfrak{g}\in G$. Fusion of symmetry defects respects the grading: the fusion of a $\mathfrak{g}_1$ defect and a $\mathfrak{g}_2$ defect results in a $\mathfrak{g}_1\mathfrak{g}_2$ defect, \ie, $\mathcal{D}_{\mathfrak{g}_1}\times \mathcal{D}_{\mathfrak{g}_2}=\mathcal{D}_{\mathfrak{g}_1\mathfrak{g}_2}$. The full topological data of a $G$-bSET is further characterized by the $G$-crossed braided tensor category $\mathcal{D}_G^\times$, which besides $\mathcal{D}_G$ also includes the symmetry action of $G$ on anyons in $\mathcal{D}$, the fractionalization of $G$ that determines the projective representation of $G$ carried by anyons, and the more abstract defectification class~\cite{BBCW,Cheng2017}. A symmetry operation permuting anyons signals non-Abelian defects in the corresponding defect sector. When the UMTC $\mathcal{D}$ is trivial, the bSET state is reduced to a bSPT. Gauging the symmetry $G$ promotes the symmetry defects to deconfined anyons and produces a new UMTC, which can be viewed as the TO of a $G$ gauge theory coupled to the original anyon system~\cite{BBCW}. 

We note that the braiding statistics of defects in the $G$-crossed theory are not gauge-invariant. This is because the defects are externally introduced instead of being deconfined excitations of the original Hamiltonian. As a result, braiding of defects generically introduces non-topological Berry phases in addition to universal statistical phases. Therefore, only the relative ratio between statistical phases of different processes is physical. The existence of these relative invariants implies a projective notion of braiding for any $G$-crossed theory, as first introduced in Refs.~\cite{You2012_projective_stats,Barkeshli2012_nematic_disloc,Barkeshli2011}.

A gapped electron system forms a fermionic TO, which is described not by a UMTC, but by a super-modular tensor category (sMTC)~\cite{Bruillard2017}. The defining feature of an sMTC is the presence of a transparent fermion $f$, which has fermionic self-statistics and braids trivially with every object in the category. Physically, $f$ represents the physical electron. It has fusion $f\times f=1$ with topological spin $1/2$. For all objects $a\in \mathcal{C}^f$, the monodromy, \ie, mutual statistics, between $a$ and $f$ is $M=1$. The simplest sMTC is $\mathrm{sVec}=\{1,f\}$, which only contains the vacuum and the physical electron. Classification of symmetry enrichment on sMTC is more complicated~\cite{Bruillard2017,Bulmash2022,Barkeshli2022,Kobayashi2022,Cheng2026}. Nevertheless, the basic defect sector viewpoint survives. After enrichment by a symmetry group $G$, one still obtains $G$-graded sectors, with the identity sector given by the original fermionic anyon theory. Familiar examples of fermionic SET states are fermionic quantum Hall states, including the Laughlin states with odd denominators, the Jain sequence~\cite{Jain1989}, the Moore--Read state~\cite{Moore1991}, and the Read--Rezayi states~\cite{Read1999}, where the fermionic TO is enriched by the charge conservation symmetry. When the underlying sMTC $\mathcal{C}^f$ is trivial, \ie, $\mathcal{C}^f=\mathrm{sVec}$, the corresponding fermionic SET (fSET) state is reduced to a fermionic SPT (fSPT) state.

Besides the fusion product of objects denoted by $\times$, we will also encounter tensor products of anyon theories. The Deligne tensor product, or stacking, of two categories is denoted by $\boxtimes$, while the tensor product of two anyon sets that need not be categories is denoted by $\otimes$. Objects in the tensor product $\mathcal{C}\boxtimes\tilde{\mathcal{C}}$ or $\mathcal{C}\otimes\tilde{\mathcal{C}}$ are labeled by composites $(a,\tilde{a})$ with $a\in\mathcal{C}$ and $\tilde{a}\in\tilde{\mathcal{C}}$. The total number of inequivalent objects in an anyon set $\mathcal{C}$ is called the rank of $\mathcal{C}$, $|\mathcal{C}|$, and hence $|\mathcal{C}\boxtimes\tilde{\mathcal{C}}|=|\mathcal{C}\otimes\tilde{\mathcal{C}}|=|\mathcal{C}||\tilde{\mathcal{C}}|$.

The only symmetry relevant to this work is charge conservation. An electron system with $U(1)$ charge conservation symmetry must satisfy the spin-charge relation. It constrains local bosons to carry even integer charge and local fermions to carry odd integer charge. In particular, the physical electron as a local fermion excitation carries unit charge $e$ under $U(1)$. The category describing this local charge assignment is $\mathrm{sRep}(U(1))$, with objects $\mathbf{m}$ labeled by integer $m$. For even or odd $m$, $\mathbf{m}$ represents a local boson or fermion, respectively. In particular, $\mathbf{1}\in\mathrm{sRep}(U(1))$ is the physical electron. Category $\mathrm{sRep}(U(1))$ can be viewed as a trivial $U(1)$-fSPT with spin-charge relation, or equivalently, a spin$_\mathbb{C}$-SPT. $U(1)$-fSET satisfying spin-charge relation is called spin$_\mathbb{C}$-SET.

\section{Classification of $2+1$D superconductors}\label{sec:classification}

SC supports vortices. For a charge-$2ne$ SC, a vortex carries flux quanta in units of $h/(2ne)$, and the $h/(2ne)$ vortex is called elementary. In algebraic language, a charge-$2ne$ SC can be viewed as a $\mathbb{Z}_{2n}$-fSET, and the vortex sector carrying flux $ph/(2ne)$ is the $p$-th $\mathbb{Z}_{2n}$ defect sector $\mathcal{C}_\mathfrak{p}$. The deconfined excitations are organized in $\mathcal{C}_\mathfrak{0}=\mathcal{C}^f$, an sMTC with physical electron as the transparent fermion $f$. Fusing two vortices yields additive flux and follows the corresponding group multiplication, \ie, $\mathcal{C}_\mathfrak{p}\times \mathcal{C}_\mathfrak{q}=\mathcal{C}_\mathfrak{p+q}$. Thus, the elementary vortex sector $\mathcal{C}_\mathfrak{1}$ with flux $h/(2ne)$ generates vortex sectors with higher flux. The most important topological data of vortices include their quantum dimensions and fusion rules~\cite{BBCW}. Topological spin, fractional charge, and modular $S$ and $T$ matrices of vortices are not gauge invariant~\cite{BBCW}. 

Given the fundamental importance of vortices, we propose a classification of SC based on its vortex sectors, especially the elementary vortex sector, that orders the intertwinement between superconductivity and topological order in the SC. Upon increasing the intertwinement, we classify SC into five categories: invertible SC, SC$^\ast$, twisted SC$^\ast$, SC$^\sharp$, and primitive SC$^\sharp$. They exhibit important properties and no-go results, which we will highlight in boxes.

\subsection{Invertible SC}\label{subsec:SC1}

We recall the classification of invertible SC for completeness. A charge-$2ne$ SC forming an invertible fermionic phase of matter is denoted as an invertible SC. For odd $n$, it has $16n$ topologically inequivalent classes labeled by $\upsilon\in\mathbb{Z}_{16n}$~\cite{16foldway,Wang20102me}. Its elementary vortex sector $\mathcal{C}_\mathfrak{1}^\mathrm{inv}$ consists of two Abelian vortices related by fusion with physical electron $f$ for even $\upsilon$, and of a single non-Abelian vortex with quantum dimension $\sqrt{2}$ invariant under fusion with $f$ for odd $\upsilon$~\cite{16foldway,Wang20102me}. On the other hand, charge-$2ne$ SC for even $n$ has only $4n$ topologically inequivalent classes, where all these classes have elementary vortex sectors $\mathcal{C}_\mathfrak{1}^\mathrm{inv}$ composed of two Abelian vortices related by $f$.

\subsection{SC$^\ast$}\label{subsec:SC2}

Invertible SC can be stacked with a charge-neutral TO, which is referred to as SC$^\ast$ in literature~\cite{Senthil2000}. Such a stacked TO must be bosonic, as a neutral fermionic TO has a charge-neutral transparent fermion that cannot be identified with the charge-$e$ physical electron due to charge mismatch. Formally, an SC$^\ast$ can be viewed as an invertible SC stacked with a UMTC $\mathcal{T}$. Then the dynamical anyon sector $\mathcal{C}^\mathrm{dym}$ in an SC$^\ast$ is identical to $\mathcal{T}$, while the elementary vortex sector is simply $\mathcal{C}_\mathfrak{1}^\mathrm{inv}\otimes\mathcal{C}^\mathrm{dym}$, \ie, elementary vortices in $\mathcal{C}_\mathfrak{1}^\mathrm{inv}$ are dressed by some deconfined anyon in $\mathcal{C}^\mathrm{dym}$. Physically, in an SC$^\ast$ the SC and the TO are topologically decoupled. The structure of its vortex sector is effectively reduced to that of the invertible SC. We also refer to this as untwisted SC$^\ast$, to distinguish from twisted SC$^\ast$ introduced below. Upon stacking specific TO, topologically inequivalent invertible SC can become equivalent SC$^\ast$~\cite{WangLevin}.

\subsection{Twisted SC$^\ast$}\label{subsec:SC3}

The untwisted SC$^\ast$ can alternately be defined via the factorization of the elementary vortex sector into $\mathcal{C}_\mathfrak{1}^\mathrm{inv}\otimes \mathcal{C}^\mathrm{dym}$. Then, there exists a choice of a reference elementary vortex whose $2n$-fold fusion contains the vacuum or physical electron. To disambiguate, here by vacuum we mean that the composite $h/e$ vortex carries no nontrivial bulk anyon. We now generalize this structure to carry a nontrivial Abelian bulk anyon. The elementary vortex sector then takes the form $\mathcal{V}_\mathfrak{1}^\mathrm{inv}\otimes\mathcal{C}^\mathrm{dym}$, where $\mathcal{V}_\mathfrak{1}^\mathrm{inv}$ is formed by fusing every vortex in $\mathcal{C}_{\mathfrak{1}}^{\mathrm{inv}}$ with the same Abelian object $v$. This $v$ satisfies the fusion rule $v^{2n}=w \in \mathcal{C}^\mathrm{dym}$, while $v$ itself does not correspond to a dynamical object or defect. The choice of $v$ is not unique, as one can redefine $v$ by fusing with an Abelian object in $\mathcal{C}_\mathfrak{0}$. If no such redefinition makes $w$ trivial, \ie, $w\neq 1$, we call the phase a {\bf twisted SC$^\ast$}. Physically, the twist implies that when there is a nontrivial $v$, the fusion result of $2n$ identical Abelian elementary vortices yields a deconfined anyon $w$, rather than the vacuum or an electron. The anyon $w$ typically signals fractionalization of $\mathbb{Z}_{2n}$ symmetry, playing the role analogous to the  ``vison'' in literature~\cite{Senthil2000,Wang2024the,Cheng2026}. In twisted SC$^\ast$, the SC and TO start to intertwine in the way that the $\mathbb{Z}_{2n}$ symmetry is typically fractionalized, making the SC a genuine $\mathbb{Z}_{2n}$-fSET.

The twisted SC$^\ast$ has the following important property:
\begin{resultbox}
A twisted SC$^\ast$ either contains non-Abelian deconfined anyons, or has condensate charge greater than $2e$. A charge-$2e$ twisted SC$^\ast$ with only Abelian deconfined anyons does not exist.
\end{resultbox}
\noindent To see this, consider a charge-$2e$ twisted SC$^\ast$ with only Abelian deconfined anyons. First note that, for any anyon $a\in \mathcal{C}^\mathrm{dym}$, its fusion with itself cannot be the physical electron $f$, as $a\times a=f$ yields $a^{-1}=a\times f$ and hence the monodromy, \ie, mutual statistics, $M_{a,f}=-1$, which violates the transparency of $f$. Therefore, $\mathcal{C}_\mathfrak{0}$ is $\mathbb{Z}_2$-graded in which $f$ is odd. Choosing $\mathcal{C}^\mathrm{dym}$ as the even component promotes $\mathcal{C}^\mathrm{dym}$ to a \emph{subcategory} of $\mathcal{C}_\mathfrak{0}$, \ie,  $\mathcal{C}_\mathfrak{0}=\mathcal{C}^\mathrm{dym}\boxtimes\mathrm{sVec}$. Upon gauging the fermion parity, the resulting UMTC can be decomposed into the product of two sub-UMTCs~\cite{Muger2003}, $\mathcal{C}^\mathrm{dym}$ and $\mathcal{E}$. Here $\mathcal{E}$ is the modular extension of $\mathrm{sVec}$, which is forced to be one class in the 16-fold-way~\cite{16foldway,Bruillard2017}. This excludes a twisted charge-$2e$ SC$^\ast$ with only Abelian deconfined anyons. See Sec.~\ref{sec:exB} for illustrative examples.

\subsection{SC$^\sharp$}\label{subsec:SC4}

If the elementary vortex sector of a SC does not admit a factorization that extracts $\mathcal{C}^\mathrm{dym}$ as a factor, we call it SC$^\sharp$. The non-factorizability definition yields two sufficient diagnostics for SC$^\sharp$:
\begin{resultbox}
A charge-$2ne$ superconductor is an SC$^\sharp$ if the rank of its elementary vortex sector is not equal to $|\mathcal{C}^\mathrm{dym}|$ or $2|\mathcal{C}^\mathrm{dym}|$ when $n$ is odd, or not equal to $2|\mathcal{C}^\mathrm{dym}|$ when $n$ is even.
\end{resultbox}
\begin{resultbox}
A charge-$2ne$ superconductor is an SC$^\sharp$ if it contains an elementary vortex whose quantum dimension is not equal to $d_a$ or $\sqrt{2}d_a$ when $n$ is odd, or not equal to $d_a$ when $n$ is even, for every deconfined anyon $a\in\mathcal{C}^\mathrm{dym}$ with quantum dimension denoted as $d_a$.
\end{resultbox}
\noindent To see this, consider a (twisted) SC$^\ast$, whose elementary vortex sector is generally $\mathcal{V}_\mathfrak{1}^\mathrm{inv}\otimes\mathcal{C}^\mathrm{dym}$. Therefore, every deconfined anyon $a\in\mathcal{C}^\mathrm{dym}$ defines an inequivalent elementary vortex $(v\times\Sigma,a)$, where $\Sigma\in \mathcal{C}_\mathfrak{1}^\mathrm{inv}$ with $|\mathcal{C}_\mathfrak{1}^\mathrm{inv}|=1,2$ for odd $n$ and $|\mathcal{C}_\mathfrak{1}^\mathrm{inv}|=2$ for even $n$. As a result, the rank of the elementary vortex sector is equal to $|\mathcal{C}^\mathrm{dym}|$ or $2|\mathcal{C}^\mathrm{dym}|$ when $n$ is odd, or equal to $2|\mathcal{C}^\mathrm{dym}|$ when $n$ is even. In addition, the quantum dimension of $(v\times\Sigma,a)$ is $d_a$ or $\sqrt{2}d_a$ for odd $n$, and $d_a$ for even $n$.

As a $\mathbb{Z}_{2n}$-fSET, the $\mathbb{Z}_{2n}$ symmetry in an SC may contribute to anyon permutation, symmetry fractionalization, or defectification~\cite{BBCW}. For a bSET, that a symmetry operation permutes anyon is equivalent to that the corresponding defect sector cannot be viewed as a trivial defect dressed by anyons. However, for SC, symmetry permuting anyon is not the only route to SC$^\sharp$. An SC$^\sharp$ can have neither anyon permutation nor symmetry fractionalization, and the nonfactorizability of the elementary vortex sector arises from fermion zero modes bound to some vortices~\cite{Aasen2021,Bulmash2022} similar to the Ising vortex in a $(p\!+\!ip)$-wave charge-$2e$ invertible SC:
\begin{resultbox}
A charge-$2ne$ superconductor is an SC$^\sharp$ if the $\mathbb{Z}_{2n}$ symmetry permutes anyons; however, SC$^\sharp$ can arise even when $\mathbb{Z}_{2n}$ contributes to none of anyon permutation, symmetry fractionalization, and defectification.
\end{resultbox}
\noindent A minimal example for illustration is the charge-$2e$ SC$^\sharp$ described by the $SO(3)_3$ sMTC. Its dynamical anyon sector is $\mathcal{C}^\mathrm{dym}=\{1,\epsilon\}$, where $\epsilon$ has topological spin $1/4$ and quantum dimension $1\!+\!\sqrt{2}$; its $h/(2e)$ vortex sector is $\mathcal{C}_\mathfrak{1}=\{\Sigma,\Sigma^f,\Pi\}$, where $\Sigma$ has quantum dimension $\sqrt{2\!+\!\sqrt{2}}$ and fusion $\Sigma\times f=\Sigma^f$, and $\Pi$ has quantum dimension $\sqrt{4\!+\!2\sqrt{2}}$ and is invariant upon fusing the physical electron $f$---this implies that $\Pi$ binds a Majorana zero mode.

The elementary vortex sector of a SC$^\sharp$ is highly constrained: 
\begin{resultbox}
 An SC$^\sharp$ cannot host an Abelian elementary vortex, or a Fibonacci elementary vortex with quantum dimension $d=(1\!+\!\sqrt{5})/2$. The minimal elementary vortex sector of an SC$^\sharp$ has quantum dimension $\sqrt{6}$, containing a single $d=\sqrt{3}$ vortex and its fusion with an electron.
\end{resultbox}
\noindent First, an SC$^\sharp$ cannot host an Abelian elementary vortex. This is because with an Abelian elementary vortex $V$, composite $(a,V)$ for each deconfined object $a\in \mathcal{C}_\mathfrak{0}$ defines an inequivalent elementary vortex, which factorizes the elementary vortex sector as $V\times \mathcal{C}_\mathfrak{0}=\mathcal{V}_\mathfrak{1}^\mathrm{inv}\otimes \mathcal{C}^\mathrm{dym}$. Fusion with an Abelian elementary vortex $V$ gives a bijection from $\mathcal{C}_\mathfrak{0}$ to $\mathcal{C}_\mathfrak{1}$. Second, an SC$^\sharp$ cannot host a Fibonacci elementary vortex with quantum dimension $(1\!+\!\sqrt{5})/2$. Suppose an SC$^\sharp$ hosts a Fibonacci elementary vortex $\Sigma$. Then its fusion with the anti-vortex yields $\Sigma\times\bar{\Sigma}=1+\varphi$, where $\varphi$ is a deconfined anyon with Fibonacci quantum dimension. According to Frobenius reciprocity, $N^\varphi_{\Sigma\bar{\Sigma}}=N^\Sigma_{\varphi\Sigma}$, it follows that $\varphi\times\Sigma=\Sigma+V$, where $V$ is an Abelian elementary vortex. Then, the elementary vortex sector can be factorized as $(V\times\mathrm{sVec})\otimes\mathcal{C}^\mathrm{dym}$. This contradicts the definition of SC$^\sharp$, as $V\times\mathrm{sVec}$ realizes a $\mathcal{V}_\mathfrak{1}^\mathrm{inv}$. Third, the minimal quantum dimension of $\mathcal{C}_\mathfrak{1}$ in an SC$^\sharp$ is $\sqrt{6}$, containing only a single $d=\sqrt{3}$ elementary vortex and its fusion with an electron~\cite{Shi2026}. This is because for a vortex with quantum dimension less than $2$, the quantum dimension must take the form of $d=2\cos{(\pi/m)}$ (integer $m\ge 3$)~\cite{Etingof2005}, and a single Ising (quantum dimension $\sqrt{2}$) or Fibonacci vortex has been shown to be excluded.

A minimal example of SC$^\sharp$ has charge-$4e$ and is described by $U(2)_{4,0}$ Chern--Simons theory obtained from quantum disordering a bilayer of $(p\!+\!ip)$-wave charge-$2e$ invertible SC by proliferating vortex--antivortex pairs between layers as a vestigial order~\cite{Shi2026} (see also Sec.~\ref{sec:exC}). It contains three Abelian neutral anyons $\{1,\,\epsilon,\,\epsilon^2\}$ where the nontrivial anyons have topological spin $1/3$, with elementary vortex sector composed of two parafermionic vortices with quantum dimension $\sqrt{3}$. 

The elementary vortex sector in an SC$^\sharp$ cannot factor out a $\mathcal{C}^\mathrm{dym}$. However, higher flux vortex sectors may still admit such a factorization. Consider a charge-$2ne$ SC$^\sharp$ whose vortex sector carrying flux $h/(ke)$ ($k>1$ and $k\mid 2n$) admits a factorization extracting $\mathcal{C}^\mathrm{dym}$. Physically, such a factorization suggests the SC$^\sharp$ be viewed as a phase where a cluster of $k$ electrons acts as an elementary particle that forms a $2n/k$-particle condensate. Only the $\mathbb{Z}_{2n/k}$ quotient of $\mathbb{Z}_{2n}$ is irreducibly intertwined with the TO. In particular, when the $h/(ke)$ sector contains a topologically trivial Abelian vortex, the $\mathbb{Z}_k$ symmetry is invisible, and the SC$^\sharp$ is effectively a $\mathbb{Z}_{2n/k}$-SET. For even $k$, it is effectively bosonic.

\subsection{Primitive SC$^\sharp$}\label{subsec:SC5}

A charge-$2ne$ SC$^\sharp$ can have the entire $\mathbb{Z}_{2n}$ symmetry irreducibly intertwined with the TO, in which no vortex sector admits a factorization extracting $\mathcal{C}^\mathrm{dym}$, and hence every vortex is non-Abelian. We term this subclass of SC$^\sharp$ as {\bf primitive SC$^\sharp$}. Interestingly, we argue, based on the topological properties of this phase alone, that it cannot arise by quantum disordering lower-charge SC via proliferation of vortex--antivortex pairs, \ie, it cannot arise as the vestigial order of lower-charge SC (see Sec.~\ref{sec:vestigial} for details and Sec.~\ref{sec:exD}--\ref{sec:exF} for examples).

\section{Unified construction from UMTC}\label{sec:construct}

\subsection{The framework}

Anyon condensation~\cite{KONG2014,Zhang2025} plays a central role in our construction of (2+1)D SC. Here anyon condensation strictly means condensing a condensable subalgebra composed of self-bosonic objects~\cite{Zhang2025}, and we only focus on the condensation of a simple current $g$. In a UMTC, an Abelian anyon $g$ with quantum dimension $d_g=1$ is called a simple current. The smallest positive integer $r$ satisfying $g^r=1$ is called the order of $g$. A simple current with order-$r$ generates a $\mathbb Z_r$ one-form symmetry of $\mathcal{D}$. The monodromy between any anyon $a$ and a simple current $g$ is $M=e^{2\pi i(s_{a\times g}-s_a-s_g)}$, which can be reduced to $M=e^{4\pi is_g}$ for $a=g$. A bosonic $g$ can be condensed directly, producing a descendant UMTC. For a fermionic $g$, one needs to stack the UMTC with $\mathrm{sVec}$, and then condense the composite $(g,f)$, which results in an sMTC. The condensation of an order-$r$ simple current $g$ in UMTC $\mathcal{D}$ has three effects. First, it grades the objects into sectors via their monodromy with $g$, \ie, charge under the $\mathbb{Z}_r$ 1-form symmetry generated by $g$. Those with trivial monodromy remain deconfined in the resulting theory. Second, it identifies objects upon fusion with $g$, \ie, $a\sim a\times g$ for all objects $a\in \mathcal{D}$, and organizes them into orbits of the $\mathbb{Z}_r$ 1-form symmetry. Third, it splits short orbits. In the simple current condensations, a length-$r$ orbit becomes a single object, while a shorter orbit of length $r^\prime$ is split into $r/r^\prime$ objects after fixed-point resolution. For fermionic simple current condensation, where the physical electron is also involved, this counting is modified by the identification with $f$; in the convention used below, an odd-length orbit splits into $r/(2r')$ objects. The fusion rules of the unsplit objects are inherited from the parent UMTC, while those of the split ones are typically determined through consistency of the fusion ring, including the associativity.

The formal classification of $\mathbb{Z}_{2n}$-fSET was proposed in Refs.~\cite{Aasen2021,Bulmash2022}. However, in the context of SC, a systematic construction and characterization of its deconfined anyon contents and vortex sectors remain unclear. In particular, known examples for SC$^\sharp$ are still scarce~\cite{Fidkowski2013,shi2025nonAbelian,Shi2026}, especially the primitive ones~\cite{Gao2026}. Here we propose a unified algebraic construction of SC, which allows us to directly obtain the deconfined anyon content and vortex sectors. It is then straightforward to apply the definition of SC$^\sharp$, the nonfactorizability of the elementary vortex sector, as its criterion. We limit the scope of construction to spinless SC formed by electrons, with no other symmetries (for example, spin, valley, lattice translation, time reversal, etc.) than electric charge conservation. Therefore, the enriching symmetry is $U(1)$ or its subgroups, where the fermion parity $(-1)^F$ is always unbroken. 

Physically, every SC has parent states that preserve $U(1)$ charge conservation symmetry. If such a parent state is gapped, it is thus described by a spin$_\mathbb{C}$-SET with spin-charge relation. As every spin$_\mathbb{C}$-SET has UMTC parents constructed by gauging the enriching symmetry~\cite{Lan2017,Bruillard2017,Galindo2025}, one may ask whether every SC can be constructed from some parent bosonic theory via anyon condensation. The answer is affirmative. To see this, consider gauging the $\mathbb{Z}_{2n}$ symmetry in the charge-$2ne$ SC, which produces a UMTC $\mathcal{D}$ viewed as a $\mathbb{Z}_{2n}$ gauge theory~\cite{BBCW,Cheng2022,Berg2026}. $\mathcal{D}$ contains a series of Abelian anyons $\mathcal{A}=\{1,c,c^2,\cdots,c^{2n-1}\}$ representing the pure gauge charges, where the topological spin of $c^p$ is $0$ or $1/2$ for even or odd integer $p$, respectively~\cite{Lan2017,Bruillard2017,Galindo2025}. Condensing $\mathcal{A}$ effectively ungauges $\mathbb{Z}_{2n}$ and returns to the $\mathbb{Z}_{2n}$-fSET describing the charge-$2ne$ SC~\cite{BBCW}. To be precise, this amounts to first stacking a trivial atomic insulator, formally denoted as $\mathrm{sRep}(U(1))$ that encodes local fermions and bosons carrying odd and even electric charge, respectively, and then condensing the condensable algebra generated by the charge-$e$ composite bosonic object $(c,\mathbf{1})$. Here $c$ labels the anyon, while $\mathbf{1}$ denotes the local charge-$e$ fermion. Generally speaking, one should stack a Chern insulator, as the gauging of the fermion parity symmetry is not unique. Nevertheless, it is always possible to choose the gauging that preserves the chiral central charge, which, at the stage of ungauging, selects the trivial atomic insulator to be stacked with. As $(c,\mathbf{1})^{2n}=(1,\mathbf{2n})$ is a charge-$2ne$ local object, condensation of $(c,\mathbf{1})$ spontaneously breaks the $U(1)$ symmetry down to $\mathbb{Z}_{2n}$~\cite{Seo2026}. The constructed charge-$2ne$ SC inherits the chiral central charge of $\mathcal{D}$, as anyon condensation does not change the chiral central charge.

When a subgroup $\mathbb{Z}_k\subset\mathbb{Z}_{2n}$ is invisible to the charge-$2ne$ SC, namely $\mathbb{Z}_k$ does not contribute to anyon permutation, symmetry fractionalization, defectification, and vortex zero mode, the SC is effectively a $\mathbb{Z}_r$-fSET with integer $r=2n/k$. Gauging the $\mathbb{Z}_r$ symmetry produces a parent UMTC, where the pure $\mathbb{Z}_r$ gauge charges are $\mathcal{A}=\{1,d,d^2,\cdots,d^{r-1}\}$. Stacking an atomic insulator to the UMTC and condensing $(d,\mathbf{k})$ returns to the SC itself. Note that, for even $k$, the fermion parity is a subgroup of $\mathbb{Z}_k$ and not gauged. Consequently, every power of $d$ is self-bosonic. For odd $k$, the fermion parity is gauged, similar to the gauging $\mathbb{Z}_{2n}$ case. This ensures $(d,\mathbf{k})$ self-bosonic and condensable. In conclusion, we have established a surjective mapping from UMTC to SC, showing that every charge-$2ne$ SC, as a $\mathbb{Z}_{2n}$-fSET with invisible $\mathbb{Z}_k\subset\mathbb{Z}_{2n}$ symmetry, can be constructed from condensing a charge-$ke$ self-bosonic object in a UMTC stacked with $\mathrm{sRep}(U(1))$. It would therefore be helpful to use the top-down perspective to construct and analyze SC starting from UMTC.

Given a UMTC $\mathcal{D}$ stacked with $\mathrm{sRep}(U(1))$, objects essential to constructing SC are the simple currents $g\in \mathcal{D}$, which are Abelian with bosonic or fermionic self statistics. It is then natural to identify an order-$r$ simple current $g$ with a cluster of $k$ electrons as long as they share the same topological spin. Physically, this amounts to condensing the charge-$ke$ bosonic object $(g,\mathbf{k})$. The lowest-charged local object in the fusion group of $(g,\mathbf{k})$ is $(1,\mathbf{2n})$ with $2n=kr$, and condensing $(g,\mathbf{k})$ gives rise to a charge-$2ne$ SC. Since $g$ generates a $\mathbb{Z}_{r=2n/k}$ one-form symmetry, the monodromy between an anyon $a\in\mathcal{D}$ and $g$ is $M=e^{2\pi i\frac{pk}{2n}}$ with $p=0,1,\cdots,2n/k\!-\!1$, which can be viewed as the charge of $a$ under the $\mathbb{Z}_{2n/k}$ one-form symmetry. Therefore, such an anyon $a$ corresponds to a vortex carrying $ph/(2ne)$ flux in the charge-$2ne$ SC from condensing $(g,\mathbf{k})$, as the Aharonov--Bohm phase for a cluster of $k$ electrons moving around a $ph/(2ne)$ vortex is exactly $\Phi=2\pi kp/(2n)$. Anyons in the UMTC with $p=0$ become the deconfined anyons in the SC. All the topological data for deconfined anyons and vortices are inherited from the parent UMTC $\mathcal{D}$. Note that, vortices carrying flux in multiples of $h/(ke)$ are trivial, in the sense that they are composed of deconfined anyons and a bare Abelian $h/(ke)$ vortex with trivial topological spin. Consequently, the constructed charge-$2ne$ SC can be viewed as a $(2n/k)$-particle condensate of charge-$ke$ elementary particles carrying trivial topology, consistent with the invisibility of $\mathbb{Z}_k$ symmetry.
\begin{resultbox}
An SC constructed from condensing $(g,\mathbf{k})$ has trivial vortex sectors with flux quanta in multiples of $h/(ke)$. For even $k$, such a phase can be viewed as a superfluid formed by charge-$ke$ bosons stacked with a trivial SC.
\end{resultbox}

The construction of SC from UMTC provides a simple but comprehensive route to read out both the deconfined anyon and the vortex sectors from the parent UMTC. It is then straightforward to classify the constructed SC by comparing its dynamical anyon sector and elementary vortex sector.

When the UMTC admits a gauge theory description, \ie, as a CS theory of a gauge group $G$ at level-$l$ denoted as $G_l$, this scheme yields an effective field theory description $[G_l\times U(1)_0]/\mathbb{Z}_{r}$ with a $U(1)$ gauge field coupled to external gauge field $A$ through a mutual CS term with coefficient $2n$. For example, the minimal charge-$4e$ SC$^\sharp$ and primitive SC$^\sharp$ are constructed by $[SU(2)_4\times U(1)_0]/\mathbb{Z}_{2}$ and $[SU(4)_4\times U(1)_0]/\mathbb{Z}_{4}$, respectively~\cite{Shi2026,Gao2026}.

\subsection{Topology and charge hierarchy}\label{sec:order}

Given a UMTC $\mathcal{D}$ stacked with an atomic insulator $\mathrm{sRep}(U(1))$, there are infinitely many types of SC that can be constructed. It is then natural to define proper hierarchies among them. Since TO and condensate charge are the principal ingredients of an SC, we define corresponding hierarchies based on topology and charge.

The topology hierarchy comes from the UMTC $\mathcal{D}$ sector. Note that, the maximal condensable algebra in $\mathcal{D}$ may not be generated by a single simple current $g$. Consider a condensable algebra $\mathcal{A}^\prime\subset\mathcal{D}\boxtimes\mathrm{sRep}(U(1))$, where $(1,\mathbf{2n})$ is the lowest charged local object in both $\mathcal{A}^\prime$ and its subalgebra $\mathcal{A}$ generated by $(g,\mathbf{k})$. Therefore, condensing $\mathcal{A}^\prime$ generates a charge-$2ne$ SC, which can also be obtained in a two-step condensation~\cite{KONG2014,DMNO,DNO}: condensing $(g,\mathbf{k})$ first to generate an intermediate charge-$2ne$ SC, and then condensing the image of $\mathcal{A}^\prime$, a charge-neutral condensable algebra, in this intermediate SC. The second-step condensation eliminates both the deconfined anyons and the vortices that are not mutually local to the image of $\mathcal{A}^\prime$, but does not break the $\mathbb{Z}_{2n}$ symmetry. Therefore, given an SC, one can successively condense the charge-neutral condensable algebra among its deconfined excitations, and finally arrive at an SC with the same physical charge but no nontrivial charge-neutral condensable algebra. Here this charge-neutral condensable algebra retains the local-charge dressing and residual symmetry structure inherited from the parent construction. This two-step condensation defines a hierarchy of SC based on their TO. It has an important consequence: condensing a neutral condensable algebra can sometimes {\em sharpen} an SC by eliminating Abelian vortices not mutually local to that algebra, \ie, it can promote (twisted) SC$^\ast$ to SC$^\sharp$, and SC$^\sharp$ to primitive SC$^\sharp$ (see Sec.~\ref{sec:exC}--\ref{sec:exD} for examples). 

The charge hierarchy arises from the local object $\mathbf{k}\in\mathrm{sRep}(U(1))$ in the construction of SC. Given an order-$r$ simple current $g\in \mathcal{D}$, each local object $\mathbf{k}$ with the same topological spin as $g$ can be paired with $g$ to form a charge-$ke$ condensable object $(g,\mathbf{k})$. This yields an infinite series of charge-$kre$ SC with increasing $k$:
\bea
\text{Fermionic }g:& \text{ charge-}re,~\text{charge-}3re,~\text{charge-}5re,\cdots.\\
\text{Bosonic }g:& \text{ charge-}2re,~\text{charge-}4re,~\text{charge-}6re,\cdots.\label{se:1}
\eea
These SC are all effectively $\mathbb{Z}_r$-SET, with $\mathbb{Z}_k=\mathbb{Z}_{2n}/\mathbb{Z}_r$ invisible. They all share the same $ph/(2ne)$ vortex sector for given $p\in\mathbb{Z}_r$, where $p=0$ represents deconfined anyons. In particular, for self-bosonic $g$, the series of SC have invisible fermion parity symmetry $(-1)^F$. Thus, these SC are essentially $\mathbb{Z}_r$-enriched {\it bosonic} topological phases formed by Cooper pairs, with trivial $h/(2e)$ vortices. Their physical electron is inherited from $(1,\mathbf{1})\in\mathrm{sRep}(U(1))$ and not identified to any anyon in $\mathcal{D}$ during the condensation. A special case is $g=1$, where the series describes charge-$2ne$ SC$^\ast$ arising from condensing a cluster of $2n$ electrons, $(1,\mathbf{2n})$. 

The anyon condensation construction yields a necessary condition for an SC$^\sharp$ being primitive. For an SC$^\sharp$, if it is constructed from condensing $(g,\mathbf{k})$ with $k>1$, its vortex sectors with flux in multiples of $h/(ke)$ are trivial. Therefore, a necessary but not sufficient condition to construct a primitive SC$^\sharp$ is:
\begin{resultbox}
A primitive SC$^\sharp$ can only be constructed by condensing $(g,\mathbf{1})$ from a parent UMTC.
\end{resultbox}
\noindent The primitive condition also imposes strong restrictions on the parent UMTC. It requires a $\mathbb{Z}_{2n}$ 1-form center symmetry generated by a fermionic simple current in the parent UMTC. In addition, the UMTC should be generally non-factorizable. To see this, consider a factorizable UMTC $\mathcal{D}=\mathcal{D}_1\boxtimes\mathcal{D}_2$, where the simple current is $(g_1,g_2)$ with $g_{1,2}\in\mathcal{D}_{1,2}$~\footnote{One caveat is that neither $g_1$ nor $g_2$ is anyonic. See Sec.~\ref{sec:exF}.}. After condensation, $(g_1,g_2)$ is identified with the physical electron. When $g_1$ itself is anyonic and not condensable, $(g_1,1)$ has nontrivial monodromy with $(g_1,g_2)$ and falls in a vortex sector. Such a vortex sector is then factorized into $[(g_1,1)\times\mathrm{sVec}]\otimes \mathcal{C}^\mathrm{dym}$, which violates the definition of primitive SC$^\sharp$. This argument can also be extended to finite quotients of factorizable UMTCs, $(\mathcal{D}_1\boxtimes\mathcal{D}_2)/\mathcal{A}$, provided $\mathcal{A}$ has trivial monodromy with both $(g_1,1)$ and $(1,g_2)$ and contains no nontrivial power of $(g_1,g_2)$. In this case, it is possible to first construct an SC from $\mathcal{D}_1\boxtimes\mathcal{D}_2$, and then perform the quotient by condensing the image of $\mathcal{A}$ to further reduce the TO of the SC, similar to the two-step condensation operation in the topological hierarchy of SC. It generally rules out parafermion-like UMTC $(G_l\boxtimes U(1)_m)/\mathcal{A}$ for $m\neq0$, with $\mathcal{A}$ generated by a composite of objects in CS theories $G_l$ and $U(1)_m$. Physically, a charge-$2ne$ primitive SC$^\sharp$ can only be constructed from a parent UMTC whose $\mathbb{Z}_{2n}$ 1-form center symmetry generated by the fermionic simple current is irreducibly intertwined with the entire UMTC.

\section{Relation to physical mechanisms}\label{sec:mechanism}

\subsection{Vestigial order and primitive SC$^\sharp$}\label{sec:vestigial}

Vestigial order remains one prominent mechanism for higher-charge SC. However, as we will argue below, a primitive SC$^\sharp$ cannot arise as a vestigial order of lower-charge SC, and thereby should be a primary order. Here we specify the definition of vestigial order as follows. Consider a charge-$re$ SC~\footnote{Here $r$ must be an even integer.} that spontaneously breaks some discrete symmetry $G$ such as time-reversal, spatial rotation, valley/layer exchange, etc, where $G$-breaking leads to $k$-fold degeneracy of the SC state manifold. A vestigial charge-$2ne$ SC with $2n=kr$ is defined as a resulting order produced by the proliferation of the vortex--anti-vortex pairs $\Sigma^{(i)}\bar{\Sigma}^{(j)}$ for all $1\le i<j\le k$ that restores the discrete $G$ symmetry. Here $\Sigma^{(i)}$ labels a specific elementary vortex $\Sigma$ in the $i$-th state of the $k$-fold degenerate manifold, and $\bar{\Sigma}^{(i)}$ is its anti-vortex. After proliferation, a single elementary vortex $\Sigma^{(i)}$ in the parent charge-$re$ SC becomes an elementary vortex in the vestigial charge-$2ne$ SC, subject to certain identification. As fusing $r$ copies of $\Sigma^{(i)}$ results in a deconfined excitation in the parent charge-$re$ SC, the $rh/(2ne)=h/(ke)$ vortex sector in the vestigial charge-$2ne$ SC is formed by a trivial $h/(ke)$ vortex dressed by a deconfined excitation. A rigorous algebraic theory of vestigial SC is left for future studies.
\begin{resultbox}
A primitive SC$^\sharp$ cannot arise from vestigial order of lower-charge SC, and should be a genuine primary order.
\end{resultbox}

The minimal charge-$4e$ example of primitive SC$^\sharp$ that cannot be obtained from a vestigial mechanism is  described by the $U(4)_{4,0}$ Chern--Simons theory~\cite{Gao2026} (see also Sec.~\ref{sec:exD}--\ref{sec:exF}).

\subsection{Anyon superconductivity}

Every SC can be constructed from condensing a composite bosonic object $(g,\mathbf{k})$ in a parent UMTC $\mathcal{D}$ stacked with an atomic insulator $\mathrm{sRep}(U(1))$, with $g\in\mathcal{D}$ and $\mathbf{k}\in\mathrm{sRep}(U(1))$. Such a $\mathcal{D}\boxtimes\mathrm{sRep}(U(1))$ does not host symmetry fractionalization. However, it does not prevent algebraic construction of SC from other parent states, for example, the fractionalized ones. In this work, we term the algebraic construction of SC from condensing fractionally charged anyons as anyon superconductivity, in parallel to the anyon superconductivity mechanism proposed for SC developed from a fractional phase~\cite{FetterHannaLaughlin,Laughlin1988,Lee1989,WilczekWittenHalperinAnyonSC} and recently generalized to SC from doping fractional quantum anomalous Hall insulators~\cite{Senthil2025,Shi2025,Pichler2025,shi2025nonAbelian,Kuhlenkamp2025,Seo2026,6bgj-bfdn,Senthil2026,Han2026,Han20261,Zhang20261,Mehta2026} and chiral spin liquids~\cite{Divic2025,Zhang20254e,Wang20266e}, or from bandwidth-tuned transitions of those fractionalized states~\cite{shi2025nonAbelian,Shi2026sc,Shi2026}.

The category $\mathcal{D}\boxtimes\mathrm{sRep}(U(1))$ employed in our construction framework can be viewed as a special type of spin$_\mathbb{C}$-SET, where the $U(1)$ symmetry is invisible to anyons. A generic spin$_\mathbb{C}$-SET, in which anyons may carry fractional electric charge, can also be used to construct SC. Consider a spin$_\mathbb{C}$-SET $\mathcal{D}^f_{U(1)}$ and a condensable algebra $\mathcal{A}\subset\mathcal{D}^f_{U(1)}$ in which the lowest charged local object is $(1,\mathbf{2n})$. Then condensing $\mathcal{A}$ spontaneously breaks the $U(1)$ symmetry of the spin$_\mathbb{C}$-SET to $\mathbb{Z}_{2n}$ and results in a charge-$2ne$ SC. Such a condensable algebra $\mathcal{A}$ can contain fractionally charged anyons. 

In $\mathcal{A}$, there always exists an Abelian object $(g,\mathbf{k})$ with charge-neutral anyon $g$ and smallest positive integer $k$, as $(1,\mathbf{2n})$ always belongs to $\mathcal{A}$. The simple current $g$ is then identified to a cluster of $k$ electrons and correspondingly determines the vortex sectors carrying flux quanta $ph/(2ne)$ with $p\in\mathbb{Z}_{2n/k}$, while the full condensable algebra $\mathcal{A}$ determines the surviving objects, identifications, and fixed-point splittings. In contrast to the construction from UMTC, here the existence of $(g,\mathbf{k})$ does not guarantee the $\mathbb{Z}_{k}$ symmetry is invisible in the constructed SC. Instead, it only implies $\mathbb{Z}_{k}$ not to permute anyons. This is because, when $U(1)$ is spontaneously broken to $\mathbb{Z}_{2n}$ via anyon condensation, there is no universal relation between $U(1)$ symmetry fractionalization class and $\mathbb{Z}_{2n}$ symmetry fractionalization class~\cite{BBCW,Aasen2021,Bulmash2022}. As a result, the $h/(ke)$ vortex sector does not have to be trivial. The precise content of the $h/(ke)$ vortex sector should be determined by the fusion rules of the other vortex sectors inherited from the parent spin$_\mathbb{C}$-SET.

The vortex sector structure also gives a useful diagnostic for when an $SC^\sharp$ phase can arise from an anyon superconductivity mechanism. In an FQH state, SC may be induced when an energetically favorable fractionally charged anyon is doped into the system~\cite{Laughlin1988,Lee1989,Senthil2025,Shi2025,shi2025nonAbelian}. Such an anyon cannot in general condense by itself, because it need not  have bosonic self statistics. In an algebraic description of anyon superconductivity, one therefore stacks an ancillary TO, which loosely speaking represents the secondary states formed by the cheap anyons, with the FQH state, and condenses a bosonic composite object formed by binding the would-be-condensed anyon to an ancillary anyon~\cite{Zhang2025,Seo2026,shi2025nonAbelian}. The role of the ancillary sector is to cancel the unwanted topological spin and make the composite condensable. Such a composite object generates a condensable algebra containing a charge-$ke$ Abelian object $[(g,a),\mathbf{k}]$, with $g$ from the FQH state and $a$ from the stacked ancillary TO, and the combined topological part $(g,a)$ carries no electric charge. The vortex sectors in the resulting SC are then graded through the monodromy with $(g,a)$. Therefore, such an SC cannot host non-Abelian vortices if both the FQH state and the stacked TO are Abelian. 
\begin{resultbox}
From an Abelian fractional phase, SC with non-Abelian vortex, including the $(p\!+\!ip)$-wave invertible charge-$2e$ SC, can arise from anyon superconductivity mechanism only when the stacked ancillary TO is non-Abelian.
\end{resultbox}
\noindent To obtain an SC$^\sharp$ from an Abelian FQH state, \ie, Laughlin state, Jain state~\cite{Jain1989}, etc, one inevitably needs ``non-minimal'' stacking of a non-Abelian TO~\cite{shi2025nonAbelian,Seo2026}. On the other hand, restricted to minimal Abelian stacking, SC$^\sharp$ can only arise from a non-Abelian FQH state, such as Moore--Read state~\cite{Moore1991}, Read--Rezayi state~\cite{Read1999}, or more general higher-Landau-level states described by non-Abelian CS theories~\cite{Wen1991}.

\section{Examples}\label{sec:ex}

In this section, we provide explicit constructions of invertible SC, SC$^\ast$, twisted SC$^\ast$, SC$^\sharp$, and primitive SC$^\sharp$. We construct all the classes of the generalized 16-fold-way invertible SC~\cite{16foldway,Wang20102me} in Sec.~\ref{sec:exA}, SC$^\ast$ and twisted SC$^\ast$ using both UMTC and anyon superconductivity in Sec.~\ref{sec:exB}, two SC$^\sharp$ series from $SU(2)_{2n}$ CS theory in Sec.~\ref{sec:exC}, charge-$2ne$ primitive SC$^\sharp$ series from $SU(2n)_{2n}$ CS theory in Sec.~\ref{sec:exD}, charge-$4e$ primitive SC$^\sharp$ series from $\mathrm{Spin}(4m\!+\!2)_{4}$ CS theory in Sec.~\ref{sec:exE}, and a charge-$6e$ primitive SC$^\sharp$ arising from B\'ezout's identity in Sec.~\ref{sec:exF}.

\subsection{Invertible SC: The generalized 16-fold way}\label{sec:exA}

As invertible fermionic topological phases without intrinsic TO, charge-$2e$ SC are classified by the 16-fold way~\cite{16foldway}, with 16 inequivalent invertible charge-$2e$ SC phases characterized by an integer $\upsilon\in\mathbb{Z}_{16}$. They can be obtained from condensing $(\psi,\mathbf{1})$ from the parent UMTC $\mathcal{D}=\mathrm{Spin(\upsilon)}_1$ stacked with an atomic insulator $\mathrm{sRep}(U(1))$, where $\psi$ is the self-fermionic anyon in $\mathrm{Spin(\upsilon)}_1$ satisfying $\psi\times\psi=1$. 

Charge-$2ne$ invertible SC are classified by $\mathbb{Z}_{16n}$ for odd $n$ and $\mathbb{Z}_{4n}$ for even $n$, respectively~\cite{Wang20102me}. In particular, for even $n$ the vortices in all the $4n$ classes are Abelian, while for odd $n$ non-Abelian vortices with quantum dimension $\sqrt{2}$ appear in half of the $16n$ classes~\cite{Wang20102me}. 

For Abelian series, we choose UMTC $\mathcal{D}=U(1)_{4n^2}^{\boxtimes \upsilon}/(\mathbb{Z}_{2n})^{\upsilon-1}$. The quotient of $(\mathbb{Z}_{2n})^{\upsilon-1}$ is imposed by condensing the following anyons: $(2n,-2n,0,\cdots,0)$, $(0,2n,-2n,0,\cdots,0)$, $\cdots$, $(0,\cdots,0,2n,-2n)$. Here the entries $2n$ and $-2n$ denotes the $2n$-th and the ($4n^2\!-\!2n$)-th anyon in the fusion group $\mathbb{Z}_{4n^2}$ of each $U(1)_{4n^2}$ sector, respectively. After quotient, $U(1)_{4n^2}^{\boxtimes \upsilon}/(\mathbb{Z}_{2n})^{\upsilon-1}$ is of rank $4n^2$, generated by two anyons $(2n,0,\cdots,0)$ and $(1,1,\cdots,1)$ with topological spin $1/2$ and $\upsilon/(8n^2)$, respectively. The charge-$2ne$ invertible SC is obtained by condensing the composite $[(2n,0,\cdots,0),\mathbf{1}]$, which divides the anyons into sectors labeled by monodromy $M=e^{2\pi i\frac{p}{2n}}$ for integer $p\in\mathbb{Z}_{2n}$. As a result, in each $ph/(2ne)$ vortex sector there are two Abelian vortices corresponding to $(p,p,\cdots,p)$ and $(p+2n,p,\cdots,p)$ related by fusing the physical electron $f\!\sim\!(2n,0,\cdots,0)$, which indeed implies an invertible SC. The vortex topological spins are $\upsilon p^2/(8n^2)$ and $\upsilon p^2/(8n^2)+(p+n)/(2n)$, consistent with those in the classification of Abelian series of invertible charge-$2ne$ SC~\cite{Wang20102me}. 

For the non-Abelian series, we choose UMTC $\mathcal{D}=[(U(1)_{4n^2}^{\boxtimes \upsilon}/(\mathbb{Z}_{2n})^{\upsilon-1})\boxtimes \mathrm{Ising}]/\mathbb{Z}_2$. The quotient of $\mathbb{Z}_2$ is implemented by condensing the composite $[(2n^2,0,\cdots,0),\psi]$, where we use $\psi$ to denote the self-fermionic anyon in the Ising category. This identification and condensation is legal only when $n$ is odd, as $(2n^2,0,\cdots,0)$ is bosonic for even $n$ while $\psi$ is always fermionic. Constraining $n$ to be odd with $\upsilon\in\mathbb{Z}_{8n}$, condensation of $[(2n,0,\cdots,0),1,\mathbf{1}]$ yields $8n$ classes of charge-$2ne$ SC hosting non-Abelian vortices, with quantum dimension $\sqrt{2}$ and topological spin $\upsilon p^2/(8n^2)+1/16$, that carry odd multiples of flux quanta $h/(2ne)$~\cite{Wang20102me}.

\subsection{(Twisted) SC$^\ast$ and anyon superconductivity}\label{sec:exB}

We use examples to construct SC$^\ast$ and twisted SC$^\ast$, respectively, from both UMTC and anyon superconductivity approaches. We perform field theory analysis to further corroborate the anyon superconductivity construction.

Consider a Laughlin-$1/8$ state of Cooper pairs with Abelian anyons forming a $\mathbb{Z}_8$ fusion group labeled by $\{1,J,J^2,\cdots,J^7\}$. $U(1)$ enrichment assigns charge $e/4$ to $J$, such that $(J,\mathbf{0})^8=(1,\mathbf{2})$. The condensable algebra is generated by the charge-$e$ bosonic object $(J^4,\mathbf{0})$, with $(J^4,\mathbf{0})^2=(1,\mathbf{2})$. Therefore, the condensation of $(J^4,\mathbf{0})$ results in a charge-$2e$ SC, with nontrivial anyon $J^2\!\sim\!J^6$ forming a residual $U(1)_2$ TO. Note that, this deconfined anyon does not inherit the $e/2$ fractional charge from the parent $U(1)_8$ state. Instead, it is charge-neutral, as the symmetry fractionalization class $H^2(\mathbb{Z}_2/(-1)^F,\mathbb{Z}_2)=H^2(\mathbb{Z}_1,\mathbb{Z}_2)=\mathbb{Z}_1$ is trivial. Consequently, this charge-$2e$ SC is an SC$^\ast$, composed of a trivial charge-$2e$ SC stacked with a neutral semion TO. In the construction from UMTC, we simply choose $\mathcal{D}=U(1)_2$ stacked with atomic insulator $\mathrm{sRep}(U(1))$ and condense $(1,\mathbf{2})$ to obtain the same SC$^\ast$. 

Such an SC$^\ast$ can be also constructed from the standard field theory approach~\cite{Laughlin1988,Lee1989,Divic2025}. Consider the following Ginzburg--Landau--CS theory,
\beq
\mathcal{L}[A]=-\frac{8}{4\pi}\alpha\mathrm{d}\alpha+\frac{2}{2\pi}\alpha\mathrm{d}A+\mathcal{L}[\phi;4\alpha],
\eeq
where boson $\phi$ sources $(J^4,\mathbf{0})$, $\alpha$ is the dynamical $U(1)$ gauge field, and $A$ is the background spin$_\mathbb{C}$ electromagnetic field. Equation of motion of $\alpha$ yields $2\pi\rho_\phi=2\nabla\times{\bm \alpha}=(1/2)\nabla\times 4{\bm \alpha}$. Thus, it is natural to have $\phi$ in a $\nu=1/2$ Laughlin state, which gives rise to the topological response,
\beq
\mathcal{L}[A]=-\frac{2}{4\pi}(\beta-2\alpha)\mathrm{d}(\beta-2\alpha)+\frac{2}{2\pi}\alpha\mathrm{d}A,\label{eq:u12}
\eeq
where $\beta$ is an emergent dynamical $U(1)$ gauge field. The anyons in the $U(1)_2$ sector are indeed charge-neutral as gauge field $(\beta-2\alpha)$ does not couple to $A$. Correspondingly, \Eq{eq:u12} describes a charge-$2e$ SC coexisting with a $U(1)_2$ UMTC, consistent with the anyon condensation analysis. This field theory construction illustrates the original anyon superconductivity mechanism proposed in Ref.~\cite{Laughlin1988}.

Stacking TO can collapse the classification of invertible SC. As an example, consider a trivial charge-$8e$ SC with trivial elementary vortex $V$. Stacking a neutral $U(1)_8$ UMTC with anyons labeled by $1,g,g^2,\cdots,g^7$ converts it to SC$^\ast$. Through relabeling $V'=g\times V$ and $g'=g^3\times f$, this SC$^\ast$ can be alternatively viewed as a nontrivial charge-$8e$ invertible SC in class $\upsilon=8$ of the $\mathbb{Z}_{16}$ classification, whose elementary vortex $V'$ has topological spin $1/16$, stacked with a neutral $U(1)_8$ UMTC generated by $g'$. Thus, upon stacking $U(1)_8$, the two inequivalent charge-$8e$ invertible SC become topologically equivalent charge-$8e$ SC$^\ast$.

It is worth noticing that coupling between dynamical and background gauge fields does not guarantee twisted SC$^\ast$ in the field theory description. Consider a charge-$2e$ SC obtained from condensing a Cooper pair $(0,\mathbf{2})$ in a Laughlin-$1/3$ state, which spontaneously breaks the $U(1)$ charge conservation symmetry to $\mathbb{Z}_2$. Its topological response theory reads,
\beq
\mathcal{L}[A]=-\frac{3}{4\pi}b\mathrm{d}b+\frac{1}{2\pi}b\mathrm{d}A+\frac{2}{2\pi}\alpha\mathrm{d}A,
\eeq
which contains a mutual CS term between $b$ and $A$. To see it is an untwisted SC$^\ast$, we note that a bare $h/e$ vortex, {\it i.e.}, a region with $\nabla\times A =2\pi$, sources an $e/3$ anyon. This means that fusing two elementary vortices induced by $\nabla\times A =\pi$ fluxes, each bounded to a $2e/3$ anyon, can result in a topologically trivial vortex with a physical electron. Such a state can be equivalently constructed from condensing $(g^6,\mathbf{1})$ in UMTC $\mathcal{D}=U(1)_{12}$ stacked with an atomic insulator. After condensation, the dynamical anyon sector $\mathcal{C}^\mathrm{dym}$ contains three Abelian anyons $1,g^2,g^4$, and the elementary vortex sector contains three Abelian vortices $g,g^3,g^5$, respectively, modulo the physical electron. Vortex $g^3$ has topological spin $3/8$ and its fusion with itself yields the physical electron, both of which are consistent with the $\upsilon=6$ class of the 16-fold-way~\cite{16foldway}. Thus, the vortex sector of this charge-$2e$ SC is factorized as $\mathcal{C}_\mathfrak{1}^\mathrm{inv}\otimes\mathcal{C}^\mathrm{dym}$, suggesting an untwisted SC$^\ast$.

A genuine twisted SC$^\ast$ requires either non-Abelian deconfined anyons, or condensate charge greater than $2e$. The simplest example illustrates the higher charge case. Consider the following Ginzburg--Landau--CS theory that describes doping charged bosons to a neutral TO,
\beq
\mathcal{L}[A]=-\frac{16}{4\pi}\alpha\mathrm{d}\alpha+\mathcal{L}[\phi;8\alpha+2A],
\eeq
where $\phi$ sources a charge-$2e$ boson $(g^8,\mathbf{2})$. Equation of motion of $\alpha$ yields $2\pi\rho_\phi=2\nabla\times{\bm \alpha}=(1/4)\nabla\times 8{\bm \alpha}$. Then it is natural to have $\phi$ in a $\nu=1/4$ Laughlin state, which gives rise to the topological response,
\beq
\mathcal{L}[A]=-\frac{16}{4\pi}\alpha\mathrm{d}\alpha-\frac{4}{4\pi}\beta\mathrm{d}\beta-\frac{1}{2\pi}\beta\mathrm{d}(8\alpha+2A).
\eeq
Defining $\tilde{\beta}=\beta+2\alpha$, we can rewrite the theory as
\beq
\mathcal{L}[A]=-\frac{4}{4\pi}\tilde{\beta}\mathrm{d}\tilde{\beta}-\frac{2}{2\pi}\tilde{\beta}\mathrm{d}A+\frac{4}{2\pi}\alpha\mathrm{d}A,
\eeq
describing a charge-$4e$ SC with a $U(1)_4$ TO. In contrast to the previous case, there is no elementary $h/(4e)$ vortex whose quadruple can fuse into a vacuum $h/e$ vortex with or without a physical electron ($4x =2$ mod $4$ has no integer solution for $x$) and therefore it is a twisted SC$^\ast$. This state can be equivalently constructed from condensing $(g^8,\mathbf{2})$ in UMTC $\mathcal{D}=U(1)_{16}$ stacked with an atomic insulator. The Abelian anyons and the elementary vortices are $1,g^2,g^4,g^6$ and $g,g^3,g^5,g^7$. Every choice of $v$ in the latter group yields $v^4=g^4$ being a nontrivial anyon. Taking $\mathcal{C}_\mathfrak{1}^\mathrm{inv}=\mathrm{sVec}$, the elementary vortex sector is thereby factorized as $\mathcal{V}_\mathfrak{1}^\mathrm{inv}\otimes\mathcal{C}^\mathrm{dym}$, suggesting a twisted SC$^\ast$.

A charge-$2e$ twisted SC$^\ast$ necessarily hosts non-Abelian deconfined anyons. Consider a parent UMTC $\mathcal{D}=\mathrm{Ising}\boxtimes U(1)_8$ stacked with an atomic insulator. Condensing $[(\psi,g^4),\mathbf{1}]$ yields a charge-$2e$ SC. The dynamical anyon sector $\mathcal{C}^\mathrm{dym}$ is formed by $(1,g^{0,2,4,6})$ and $(\sigma,g^{1,3})$, where $\sigma$ is the Ising anyon~\cite{Aasen2021}. Note that, here $\mathcal{C}^\mathrm{dym}$ is not a category as the fusion is not closed. The elementary vortex sector is $\mathcal{V}_\mathfrak{1}^\mathrm{inv}\otimes\mathcal{C}^\mathrm{dym}$ with $v=(1,g)$. The fusion result $v\times v=(1,g^2)$ cannot be trivialized by redefining $v$ via attaching any Abelian anyon, which implies a twisted SC$^\ast$. This state can be alternatively viewed as stacking a trivial charge-$2e$ SC to a Pfaffian quantum Hall state, where the fractionalization of $U(1)$ charge symmetry in the Pfaffian state results in twisted SC$^\ast$. In other words, stacking an invertible SC on a spin$_\mathbb{C}$-SET with nontrivial fractionalization could result in twisted SC$^\ast$. Only when the topological order is neutral does this necessarily result in an untwisted SC$^\ast$. This can be further seen from the field theory description~\cite{Seiberg2016,Ma2020,Zou2020,Shi2025},
\bea
\mathcal{L}[A]=&-\frac{2}{4\pi}\trace{\alpha\mathrm{d}\alpha+\frac{2}{3}\alpha^3}+\frac{3}{4\pi}(\mathrm{tr}\,\alpha)\mathrm{d}(\mathrm{tr}\,\alpha)\\
&+\frac{1}{2\pi}(\mathrm{tr}\,\alpha)\mathrm{d}A+\frac{2}{2\pi}\beta\mathrm{d}A,
\eea
where the $U(2)$ gauge field $\alpha$ describes the Pfaffian state, and the $U(1)$ gauge field $\beta$ suggests a charge-$2e$ SC. From the field theory, an $h/e$ vortex fused from two $h/(2e)$ vortices has unit charge under $\trace{\alpha}$, {\it i.e.}, bound to a $(1,g^2)$ anyon. Similar to the charge-$4e$ twisted SC$^\ast$ discussed above, it is impossible to trivialize this bound $(1,g^2)$ anyon by attaching deconfined excitations to the $h/(2e)$ vortex. This exemplifies a minimal charge-$2e$ twisted SC$^\ast$ state.

\subsection{Two SC$^\sharp$ series from $SU(2)_{2n}$}\label{sec:exC}

$SU(2)$ CS theory has been widely employed to describe SC with non-Abelian vortices~\cite{Ma2020,Zou2020,Shi2025,Gao2026,shi2025nonAbelian,Shi2026,Wang20266e}. In particular, $U(2)_{2,0}=[SU(2)_2\times U(1)_0]/\mathbb{Z}_2$ describes a charge-$2e$ invertible SC hosting an Ising vortex~\cite{Seiberg2016,Ma2020,Zou2020,Shi2025}. Here we mainly focus on $n>1$ cases. The cases of negative $n$ can be obtained by taking time-reversal conjugation. The $SU(2)_{2n}$ CS theory has $(2n\!+\!1)$ anyons labeled by the $SU(2)$ spin $j=0,1/2,1,\cdots,n$, where the $j\!=\!n$ anyon is an order-$2$ simple current with topological spin $n/2$. It is self-fermionic for odd $n$ and bosonic for even $n$. After stacking with an atomic insulator $\mathrm{sRep}(U(1))$, the composite $(j\!=\!n,\mathbf{n})$ is automatically bosonic. As $(j\!=\!n,\mathbf{n})^2=(j\!=\!0,\mathbf{2n})$ is a local charge-$2ne$ boson, condensation of $(j\!=\!n,\mathbf{n})$ generates a charge-$2ne$ SC, with $j\!=\!n$ identified as a cluster of $n$ electrons. During the condensation, anyons with integer $j$ remain deconfined with identification $j\sim(n-j)$ and contribute to the dynamical anyon sector of the constructed SC. Anyons with half-odd $j$ have monodromy $M=-1$ between the $j\!=\!n$ anyon, and form the elementary vortex sector.

SC constructed from condensing $(j\!=\!n,\mathbf{n})$ exhibit an even--odd effect. With even $n$, $j\!=\!n$ and $\mathbf{n}$ are both bosonic, suggesting the charge-$2ne$ SC being effectively bosonic. Here $j\!=\!n/2$ fixed under the $\mathbb{Z}_2$ 1-form symmetry generated by $j\!=\!n$ falls in the dynamical anyon sector and splits into two deconfined anyons. Therefore, the rank of the dynamical anyon sector is $(n\!+\!4)/2$, while the elementary vortex sector has rank $n$. For odd $n$, the rank of the elementary vortex sector is equal to neither one nor two times that of the dynamical anyon sector; for even $n$, the rank of the elementary vortex sector is not equal to $2$ times that of the dynamical anyon sector. These suggest the corresponding charge-$2ne$ SC being SC$^\sharp$. 

With odd $n$, $j\!=\!n$ and $\mathbf{n}$ are both fermionic, suggesting an intrinsically fermionic phase after condensation. Now $j\!=\!n/2$ fixed under the $\mathbb{Z}_2$ 1-form symmetry generated by $j\!=\!n$ falls in the elementary vortex sector. However, because the condensed current is identified with the physical fermion rather than with the vacuum, the fixed-point resolution differs from ordinary bosonic condensation. As physical electron is charged and hence not fixed under the fermion parity symmetry, $j\!=\!n/2$ does not split during the condensation of $(j\!=\!n,\mathbf{n})$. Instead, it binds a Majorana zero mode and returns to itself upon fusing an electron, similar to the vortex in the $(p\!+\!ip)$-wave charge-$2e$ SC. Consequently, the dynamical anyon sector and the elementary vortex sector have rank $(n+1)/2$ and $n$, respectively. For $n>1$, the corresponding charge-$2ne$ SC is an SC$^\sharp$. 

In summary, condensing $(j\!=\!n,\mathbf{n})$ in $SU(2)_{2n}$ UMTC stacked with $\mathrm{sRep}(U(1))$ generates a series of charge-$2ne$ SC$^\sharp$ for $n>1$. The topological quantum field theory describing these SC$^\sharp$ is a $U(2)_{2n,0}$ CS theory,
\bea
\mathcal{L}[A]=&-\frac{2n}{4\pi}\trace{\alpha\mathrm{d}\alpha+\frac{2}{3}\alpha^3}+\frac{n}{4\pi}(\mathrm{tr}\,\alpha)\mathrm{d}(\mathrm{tr}\,\alpha)\\
&-\frac{n}{2\pi}(\mathrm{tr}\,\alpha)\mathrm{d}A,
\eea
where $\alpha$ is a $U(2)$ gauge field. These charge-$2ne$ SC$^\sharp$ states have been constructed from anyon superconductivity mechanism via field theory in doped fractional quantum Hall states~\cite{shi2025nonAbelian} and chiral spin liquids~\cite{Wang20266e}. We note that, the charge-$4e$ SC$^\sharp$ in this series, hosting $\mathbb{Z}_3$ Abelian anyons in the bulk and two elementary vortices with quantum dimension $\sqrt{3}$~\cite{Shi2026}, is the SC$^\sharp$ with smallest total quantum dimension.

This series provides a simple example to illustrate topology hierarchy that sharpens SC. Consider $\mathcal{D}$ composed of stacking $SU(2)_4$ to a toric code UMTC whose nontrivial anyons are denoted as $e_\mathrm{TC},m_\mathrm{TC},\psi_\mathrm{TC}$. Condensation of both $[(j\!=\!0,e_\mathrm{TC}),\mathbf{0}]$ and $[(j\!=\!2,e_\mathrm{TC}),\mathbf{2}]$ produces the above charge-$4e$ SC$^\sharp$, while condensing only the latter yields a charge-$4e$ SC$^\ast$ with $\mathcal{C}^\mathrm{dym}=\{(j\!=\!0,1),(j\!=\!1,1),(j\!=\!2,1),(j\!=\!1/2,m_\mathrm{TC}),(j\!=\!3/2,m_\mathrm{TC})\}$ and $\mathcal{C}_\mathfrak{1}=\{(j\!=\!0,m_\mathrm{TC}),(j\!=\!1,m_\mathrm{TC}),(j\!=\!2,m_\mathrm{TC}),(j\!=\!1/2,1),(j\!=\!3/2,1)\}$. Starting from this SC$^\ast$, further condensing the image of $[(j\!=\!0,e_\mathrm{TC}),\mathbf{0}]\sim [(j\!=\!2,1),\mathbf{2}]$ eliminates the Abelian vortices and reproduces the SC$^\sharp$~\footnote{Here the monodromy includes the $-1$ sign between local charge $\mathbf{2}$ and $h/(4e)$ flux quanta.}.

According to the charge hierarchy, it is always possible to condense $(j\!=\!n,\mathbf{2})$ or $(j\!=\!n,\mathbf{1})$ in $SU(2)_{2n}$ CS theory with even or odd $n$, respectively. This defines a series of charge-$2le$ ($l=\mathrm{gcd}(n,2)$) SC$^\sharp$ for $n>1$, described by,
\bea
\mathcal{L}[A]=&-\frac{2n}{4\pi}\trace{\alpha\mathrm{d}\alpha+\frac{2}{3}\alpha^3}+\frac{n}{4\pi}(\mathrm{tr}\,\alpha)\mathrm{d}(\mathrm{tr}\,\alpha)\\
&-\frac{\mathrm{gcd}(n,2)}{2\pi}(\mathrm{tr}\,\alpha)\mathrm{d}A.
\eea
One particular example is for $n=3$. Condensing $(j\!=\!3,\mathbf{1})$ yields a charge-$2e$ SC$^\sharp$ described by $SO(3)_3$ sMTC. It has only one nontrivial anyon $\epsilon$ arising from $(j\!=\!1)\sim(j\!=\!2)$ with topological spin $1/4$ and quantum dimension $1\!+\!\sqrt{2}$. The $h/(2e)$ vortex sector is composed of three non-Abelian vortices $\Sigma$, $\Pi$, and $\Sigma^f$ with quantum dimension $\sqrt{2\!+\!\sqrt{2}}$, $\sqrt{4\!+\!2\sqrt{2}}$, and $\sqrt{2\!+\!\sqrt{2}}$, arising from $j\!=\!1/2$, $j\!=\!3/2$, and $j\!=\!5/2$, respectively. The $\Sigma^f$ vortex is obtained by fusing $\Sigma$ with the physical electron $f$, while the $\Pi$ vortex is invariant upon fusing $f$ and binds a Majorana zero mode. This charge-$2e$ SC$^\sharp$ can be realized as the surface state of certain 3D topological SC~\cite{Fidkowski2013}.

\subsection{Charge-$2ne$ primitive SC$^\sharp$ series from $SU(2n)_{2n}$}\label{sec:exD}

Primitive SC$^\sharp$ with arbitrary condensate charge is rare. Here we propose that primitive SC$^\sharp$ can arise from condensing $(\Lambda,\mathbf{1})$ from $SU(2n)_{2n}$ CS theory ($n>1$) stacked with an atomic insulator~\cite{Gao2026}, where $\Lambda=(2n,0,\cdots,0)$ is the fermionic simple current generating a $\mathbb{Z}_{2n}$ 1-form symmetry in $SU(2n)_{2n}$. The corresponding topological quantum field theory describing such SC is $U(2n)_{2n,0}$~\cite{Gao2026},
\bea
\mathcal{L}[A]=&-\frac{2n}{4\pi}\trace{\alpha\mathrm{d}\alpha+\frac{2}{3}\alpha^3}+\frac{1}{4\pi}(\mathrm{tr}\,\alpha)\mathrm{d}(\mathrm{tr}\,\alpha)\\
&-\frac{1}{2\pi}(\mathrm{tr}\,\alpha)\mathrm{d}A,
\eea
with $U(2n)$ gauge field $\alpha$. While a rigorous proof is deferred for future studies, we provide two cases for $n=2$ and $n=3$ for illustration.

Anyons in $SU(2n)_{2n}$ CS theory are labeled by their $SU(2n)$ weight $(\lambda_1,\lambda_2,\cdots,\lambda_{2n-1})$, with nonnegative integers $\lambda_i$ satisfying $\sum_i\lambda_i\le 2n$. The fermionic simple current is $\Lambda=(2n,0,\cdots,0)$ with topological spin $(n\!-\!1/2)$. The fusion of an anyon in $SU(2n)_{2n}$ with $\Lambda$ is $\Lambda\times (\lambda_1,\lambda_2,\lambda_3,\cdots,\lambda_{2n-1})=(2n-\sum_{i=1}^{2n-1}\lambda_i,\lambda_1,\lambda_2,\cdots,\lambda_{2n-2})$. The monodromy between an anyon and $\Lambda$ is $M=e^{2\pi iQ/(2n)}$, where $Q=\sum_i i\lambda_i$. Consequently, for $n=2$, condensing $[(4,0,0),\mathbf{1}]$ in $SU(4)_4\boxtimes\mathrm{sRep}(U(1))$ results in the following sectors (physical electron omitted).
\begin{itemize}
    \item The dynamical anyon sector:
    \bea
    & 1\sim (0,0,0)\sim (4,0,0)\sim (0,4,0)\sim (0,0,4),\\
    & \epsilon\sim (0,1,2)\sim (1,0,1)\sim (2,1,0)\sim (1,2,1),\\
    & \tau_\pm \sim (0,2,0)\sim (2,0,2),
    \eea
    where the last length-2 orbit splits into two anyons $\tau_\pm$. Each anyon can be fused with the physical electron to form $f$, $\epsilon^f$, or $\tau_\pm^f$. The $\epsilon$ anyon has topological spin $1/2$ and quantum dimension $(3\!+\!2\sqrt{2})$, while the $\tau_\pm$ anyon has topological spin $1/4$ and quantum dimension $(1\!+\!\sqrt{2})$.
    \item The $h/(4e)$ vortex sector:
    \bea
    \Sigma_1\sim (0,0,3)\sim (1,0,0)\sim (3,1,0)\sim (0,3,1),\\
    \Sigma_2\sim (0,1,1)\sim (2,0,1)\sim (1,2,0)\sim (1,1,2),
    \eea
    with quantum dimension $\sqrt{4\!+\!2\sqrt{2}}$ and $(1\!+\!\sqrt{2})\sqrt{4\!+\!2\sqrt{2}}$, respectively. Each vortex can be fused with the physical electron to form $\Sigma_{1,2}^f$.
    \item The $h/(2e)$ vortex sector:
    \bea
    & \Gamma_1\sim (0,0,2)\sim (2,0,0)\sim (2,2,0)\sim (0,2,2),\\
    & \Gamma_2\sim (0,1,0)\sim (3,0,1)\sim (0,3,0)\sim (1,0,3),\\
    & \Pi_\pm \sim (1,1,1),
    \eea
    where the last length-1 orbit fixed under the $\mathbb{Z}_4$ 1-form symmetry splits into two vortices $\Pi_\pm$, instead of four, as the physical electron is not fixed under $(-1)^F$. The $\Gamma_{1,2}$ vortices have quantum dimension $(2\!+\!\sqrt{2})$, and form $\Gamma_{1,2}^f$ when fused with the physical electron $f$. The $\Pi_\pm$ vortices have quantum dimension $(2\!+\!2\sqrt{2})$ and are invariant upon fusing $f$, \ie, $f\times \Pi_\pm=\Pi_\pm$.
    \item The $3h/(4e)$ vortex sector:
    \bea
    \bar{\Sigma}_1\sim (0,0,1)\sim (3,0,0)\sim (1,3,0)\sim (0,1,3),\\
    \bar{\Sigma}_2\sim (0,2,1)\sim (1,0,2)\sim (1,1,0)\sim (2,1,1),
    \eea
    with quantum dimension $\sqrt{4\!+\!2\sqrt{2}}$ and $(1\!+\!\sqrt{2})\sqrt{4\!+\!2\sqrt{2}}$, respectively. Each vortex can be fused with the physical electron to form $\bar{\Sigma}_{1,2}^f$.
\end{itemize}
Clearly, this charge-$4e$ SC is indeed an SC$^\sharp$. The fusion rules of the unsplit anyons and vortices are directly inherited from those in the parent $SU(4)_4$. For the split ones, their fusion rules are solved using the associativity of fusion recursively and symmetry upon fusing an electron. The complete fusion rules are listed in Appendix~\ref{app:fusion}. To our knowledge, this state is already the charge-$4e$ primitive SC$^\sharp$ with smallest total quantum dimension.

If one instead condenses the charge-$2e$ composite $[(0,4,0),\mathbf{2}]$ built from the order-2 boson $(0,4,0)$ in $SU(4)_4\boxtimes\mathrm{sRep}(U(1))$, the resulting charge-$4e$ state remains an SC$^\sharp$ but is no longer primitive. This is because the $h/(2e)$ sector now can be factorized as $\mathcal{C}_\mathfrak{2}=\mathcal{C}^\mathrm{inv}_\mathfrak{2}\otimes\mathcal{C}^\mathrm{dym}$. Further condensing the image of $[(4,0,0),\mathbf{1}]$ in this charge-$4e$ SC$^\sharp$ eliminates the Abelian vortex in $\mathcal{C}_\mathfrak{2}$ and promotes the state to the charge-$4e$ primitive SC$^\sharp$ described by $U(4)_{4,0}$ CS theory.

The same analysis extends to $SU(6)_6$, where the condensation of $[(6,0,0,0,0),\mathbf{1}]$ yields a charge-$6e$ SC. The rank of the dynamical anyon sector is $16$, while the ranks of the $h/(6e)$, $h/(3e)$, and $h/(2e)$ vortex sectors are $25$, $26$, and $28$, respectively, including the physical electron. In particular, there is one and four vortices invariant upon fusing an electron in the $h/(6e)$ and $h/(2e)$ vortex sectors, respectively. Clearly, this charge-$6e$ SC is a primitive SC$^\sharp$.


\subsection{Charge-$4e$ primitive SC$^\sharp$ series from $\mathrm{Spin}(4m\!+\!2)_4$}\label{sec:exE}

A charge-$2ne$ primitive SC$^\sharp$ implies a parent UMTC with an order-$2n$ fermionic simple current, which significantly limits the available examples of such an SC$^\sharp$ for generic $n$. For $n=2$, however, there are more choices besides $SU(4)_4$ of the parent UMTC such as the $\mathrm{Spin}(4m\!+\!2)_4$ CS theory. We propose that $\mathrm{Spin}(4m\!+\!2)_4$ CS theory provides parent UMTCs for a series of charge-$4e$ primitive SC$^\sharp$. In particular, $\mathrm{Spin}(6)_4$ coincides with $SU(4)_4$ described in the above subsection, as $\mathrm{Spin}(6)=SU(4)$. The minimal distinct example is $\mathrm{Spin}(10)_4$. The ranks of the dynamical anyon sector, the $h/(4e)$, the $h/(2e)$, and the $3h/(4e)$ vortex sectors are $9$, $6$, $14$, and $6$, respectively, which indeed supports a charge-$4e$ primitive SC. Detailed operator contents and fusion rules are given in Appendix~\ref{app:fusion2}. 

\subsection{Charge-$6e$ primitive SC$^\sharp$ from B\'ezout's identity}\label{sec:exF}

As discussed in Sec.~\ref{sec:order}, a factorizable UMTC $\mathcal{D}=\mathcal{D}_1\boxtimes\mathcal{D}_2$ with condensable simple current $(g_1,g_2)$ cannot yield a primitive SC$^\sharp$ when $(g_1,1)$ (or equivalently $(1,g_2)$) has nontrivial monodromy with $(g_1,g_2)$. That is because $[(g_1,1)\times\mathrm{sVec}]\otimes\mathcal{C}^\mathrm{dym}$ generates a vortex sector in the constructed SC. However, when both $(g_1,1)$ and $(1,g_2)$ have trivial monodromy with $(g_1,g_2)$, namely they are self-bosonic or fermionic, these two anyons belong to the dynamical anyon sector in the constructed SC. This implies such an SC to be potentially primitive. Without losing generality, we assume $g_1$ with order $r_1$ to be fermionic, $g_2$ with order $r_2$ to be bosonic, and $r_1$, $r_2$ coprime. The condensable algebra generated by $[(g_1,g_2),\mathbf{1}]$ can be thereby viewed as generated by $[(g_1,1),\mathbf{r}_2]$ and $[(1,g_2),\mathbf{r}_1]$ together, up to relabeling the anyons. Condensing $(g_1,\mathbf{r}_2)$ in $\mathcal{D}_1$ and $(g_2,\mathbf{r}_1)$ in $\mathcal{D}_2$ both yield a charge-$r_1r_2e$ SC, where the trivial vortex sectors have flux quanta in multiples of $h/(r_2e)$ and $h/(r_1e)$ in the former and the latter one, respectively. Products of their nontrivial vortex sectors reproduce every vortex sector in the SC constructed from condensing $[(g_1,g_2),\mathbf{1}]$, according to B\'ezout's identity: {\it Equation $r_1x+r_2y=1$ has integer solutions for $x$ and $y$ when $r_1$ and $r_2$ are coprime}. This provides an alternative way to construct charge-$2ne$ primitive SC$^\sharp$ with $n>2$, provided order-$2n$ fermionic simple current with $n>2$ is rare in non-factorizable UMTC. In particular, the coprime condition of orders $r_1$ and $r_2$ constrains that the charge $2n\neq 2^l$ for positive integer $l$. This rules out $4e$ and $8e$ for low condensate charge series, and may suggest that charge-$6e$ primitive SC$^\sharp$ be more achievable than charge-$4e$ ones.

As an example, we consider a $Sp(4)_3\boxtimes SU(3)_3$ CS theory~\footnote{The simpler $SU(2)_2\boxtimes SU(3)_3$ produces a charge-$6e$ SC$^\sharp$ with $h/(2e)$ vortex sector factorized as $\sigma\times\mathcal{C}^\mathrm{dym}$, where $\sigma$ is the Ising anyon in $SU(2)_2$.}. We use $Sp(4)_3$ and $SU(3)_3$ weights $(\lambda_1,\lambda_2)$ with $\lambda_1+\lambda_2\le 3$ to label the anyons. CS theory $Sp(4)_3$ has an order-2 anyon $g_1=(0,3)$. Condensing $[(0,3),\mathbf{3}]$ in $Sp(4)_3$ yields a charge-$6e$ SC$^\sharp$. The dynamical anyon sector realizes the $SO(5)_3$ spin CS theory, which has three anyons $1$, $\epsilon_1$, $\epsilon_2$ with topological spin $0$, $1/3$, $1/2$, and quantum dimension $1$, $(1\!+\!\sqrt{3})$, $(2\!+\!\sqrt{3})$, respectively. Each anyon can be fused with the physical electron $f$. The elementary $h/(6e)$ vortex sector has three vortices $\Gamma$, $\Pi_1$, $\Pi_2$ with quantum dimension $(1\!+\!\sqrt{3})$, $(1\!+\!\sqrt{3})$, $(3\!+\!\sqrt{3})$, respectively. In particular, $\Gamma$ can be fused with $f$ to form $\Gamma^f$, while $\Pi_{1,2}$ are invariant upon fusing $f$. The $h/(3e)$ vortex sector is trivial. On the other hand, $SU(3)_3$ has an order-$3$ anyon $g_2=(0,3)$, and condensing $[(0,3),\mathbf{2}]$ in $SU(3)_3$ yields another charge-$6e$ SC$^\sharp$. The dynamical anyon sector reproduces the $SO(8)_1$ theory with three fermionic Abelian anyons $\zeta_{1,2,3}$ forming a $\mathbb{Z}_2\times\mathbb{Z}_2$ fusion group, each of which can be fused with $f$ to form $\zeta_{1,2,3}^f$. The $h/(6e)$ and $h/(3e)$ vortex sector contains one vortex $\Sigma$ and $\bar{\Sigma}$, respectively, with quantum dimension $2$. Fusion with $f$ produces $\Sigma^f$ and $\bar{\Sigma}^f$. The $h/(2e)$ vortex is trivial. 

Taken together, the sectors of the ``B\'ezout'' charge-$6e$ SC$^\sharp$ from condensing $\{[(0,3),(0,3)],\mathbf{1}\}$ are organized as follows. The dynamical anyon sector is $SO(5)_3\boxtimes SO(8)_1=\{1,\epsilon_i,\zeta_j,\epsilon_i\times\zeta_j\}$. The $h/(6e)$ elementary vortex sector is $\{\sigma\times\Gamma,\sigma\times\Pi_i\}$, where $\sigma$ is the ``anyon part'' of $\Sigma$, \ie, $\Sigma$ fused with a trivial $-h/(6e)$ vortex. The $h/(3e)$ vortex sector is $\{\bar{\Sigma},\epsilon_i\times \bar{\Sigma}\}$. The $h/(2e)$ vortex sector is $\{\Gamma,\Pi_i,\zeta_j\times\Gamma,\zeta_j\times\Pi_i\}$. The $2h/(3e)$ and $5h/(6e)$ vortex sectors contain anti-vortices of the $h/(3e)$ and $h/(6e)$ sectors, respectively. By definition, this charge-$6e$ SC$^\sharp$ is indeed primitive.

\section{Conclusion and outlook}\label{sec:outlook}

In this work we developed a vortex sector framework for superconductors with intrinsic topological order. Since a charge-$2ne$ superconductor breaks $U(1)$ charge conservation to $\mathbb{Z}_{2n}$, its superconducting vortices are naturally viewed as $\mathbb{Z}_{2n}$ defects. The elementary $h/(2ne)$ vortex sector is therefore a basic piece of topological data of the superconducting phase. We showed that this sector provides a sharp way to distinguish superconducting phases in which superconductivity and TO simply coexist from those in which they are irreducibly intertwined.

This viewpoint led us to organize superconductors into invertible SC, untwisted and twisted SC$^\ast$, and SC$^\sharp$ phases. An invertible SC phase has each vortex sector composed of either two Abelian vortices related by fusing an electron, or a non-Abelian Ising vortex invariant upon such a fusion. In an SC$^\ast$ or twisted SC$^\ast$ phase, the elementary vortex sector factorizes into an invertible superconductor vortex sector dressed by deconfined anyons, possibly with a twist. By contrast, an SC$^\sharp$ phase is defined by the failure of this factorization. In such a phase, the non-Abelian-ness of the elementary vortex is not an optional dressing by hidden topological order; it is intrinsic to the topologically ordered superconductor itself. This nonfactorization gives practical rank and quantum dimension diagnostics for identifying SC$^\sharp$ from its deconfined anyon and elementary vortex data.

SC$^\sharp$ not only defines a new class of SC, but also provides a platform to visualize and manipulate non-Abelian vortex. In contrast to SC$^\ast$, different elementary vortices in SC$^\sharp$ are on equal footing. Non-Abelian vortex with quantum dimension $d>\sqrt{2}$ in SC$^\sharp$ may also contribute to universal quantum computing when braiding is supplemented by suitable interferometric measurements~\cite{Shi2026}, which is believed not to be achievable from Ising vortex alone.

We also presented a unified anyon condensation construction of superconductors. Starting from a parent UMTC stacked with an atomic insulator $\mathrm{sRep}(U(1))$, condensing a charged bosonic simple current composite produces a superconducting phase whose deconfined anyons and vortex sectors are determined by the monodromy with the condensed simple current. This turns the characterization of superconducting vortices into a computable problem in the parent UMTC. The same framework also clarifies the relation to anyon superconductivity mechanism achieved by condensing fractionally charged anyons in a spin$_\mathbb{C}$-SET. When such condensation produces superconductivity, the deconfined anyons and vortex sectors are determined by the monodromy with the simple current in the condensable algebra, instead of the condensed fractional-charge anyon itself.

We illustrated the framework with several classes of examples. The construction reproduces the generalized 16-fold way superconductors and SC$^\ast$ phases, and produces SC$^\sharp$ series from Chern--Simons theories such as $SU(2)_{2n}$. We further identified primitive SC$^\sharp$,
for which all nontrivial vortex sectors exhibit nonfactorization,
not only the elementary one. Primitive SC$^\sharp$ are argued to be primary orders, not able to arise as vestigial orders, \ie, by disordering relative phases of lower-charge condensates. Examples include proposed series from $SU(2n)_{2n}$ and $\mathrm{Spin}(4m\!+\!2)_4$, as well as a charge-$6e$ example by applying B\'ezout's identity. These examples show that nonfactorizable vortex sector structure is not an isolated phenomenon, but a robust organizing principle for higher-charge topological superconductivity.

Several directions remain open. Establishing the proposed primitive series in full generality and developing systematic methods for fixed-point resolution would extend the explicit constructions presented here. An algebraic theory of vortex--anti-vortex proliferation would also be valuable, which should determine the deconfined anyons and vortex sectors of a vestigial superconductor from the topological data of its parent lower-charge superconductor, and sharpen the proposed obstruction underlying primitive SC$^\sharp$. Extending the current algebraic framework to incorporate spin, crystalline, valley, or time-reversal symmetries may further broaden the landscape of topological charge-$2ne$ superconductors. More broadly, our results motivate asking whether an analogous notion of SSB$^\sharp$ can be formulated for other symmetry-breaking phases with intrinsic topological order. The central question is whether their defect sectors admit a separation into an invertible order contribution and a deconfined anyon contribution, or instead exhibit an irreducible intertwinement analogous to SC$^\sharp$. Such a perspective would extend the distinction between coexistence and intertwinement beyond superconductivity, identifying the topological structure of defects as an essential part of characterizing symmetry-breaking order.

\section{Acknowledgments}
We acknowledge Meng Cheng, Chao-Ming Jian, Chunxiao Liu, Srinivas Raghu, Taige Wang, and Zhengyan Darius Shi for helpful discussions. We are coordinating submission with a related work~\cite{Toappear1}, which discusses topological bosonic superfluids. Z.-Q.~G. is supported by the Berkeley graduate program. Y.-Q.~W. was supported by the JQI postdoctoral fellowship at the University of Maryland. Z.~H. is supported by the Gordon and Betty Moore Foundation EPiQS Award 8683. Z.~H. and A.~V. are supported by the Simons Collaboration on Ultra-Quantum Matter, which is a grant from the Simons Foundation (651440, A.~V.). We used ChatGPT 5.6 Pro and 6 Astra to assist the derivations in the appendices and check the proofs in the main text and to improve clarity of parts of the text.

\bibliographystyle{apsrev4-2}

\bibliography{2ne.bib}


\clearpage

\appendix

\section{Fusion rules of charge-$4e$ primitive SC$^\sharp$ from $SU(4)_4$}\label{app:fusion}

\bea
&\epsilon\times\epsilon
=
1+2\epsilon+2\epsilon^f
+\tau_++\tau_-+\tau_+^f+\tau_-^f,\\
&\epsilon\times\tau_\pm
=
\tau_\mp+\epsilon+\epsilon^f,\\
&\tau_\pm\times\tau_\pm
=
1+\tau_\pm+\tau_\pm^f,\\
&\tau_+\times\tau_-=\epsilon.
\eea

\bea
&\epsilon\times\Sigma_1
=
\Sigma_1+\Sigma_2+\Sigma_2^f,\\
&\epsilon\times\Sigma_2
=
\Sigma_1+\Sigma_1^f+3\Sigma_2+2\Sigma_2^f,\\
&\tau_\pm\times\Sigma_1=\Sigma_2,\\
&\tau_\pm\times\Sigma_2
=
\Sigma_1+\Sigma_2+\Sigma_2^f.
\eea

\bea
&\epsilon\times\Gamma_1
=
\Gamma_1+\Gamma_2+\Gamma_2^f+\Pi_++\Pi_-,\\
&\epsilon\times\Gamma_2
=
\Gamma_1+\Gamma_1^f+\Gamma_2+\Pi_++\Pi_-,\\
&\tau_\pm\times\Gamma_1
=
\Gamma_1^f+\Pi_\pm,\\
&\tau_\pm\times\Gamma_2
=
\Gamma_2+\Pi_\mp,\\
&\epsilon\times\Pi_\pm
=
\Pi_\mp+\Pi_++\Pi_-
+\Gamma_1+\Gamma_1^f+\Gamma_2+\Gamma_2^f,\\
&\tau_\pm\times\Pi_\pm
=
\Pi_\pm+\Gamma_1+\Gamma_1^f,\\
&\tau_\pm\times\Pi_\mp
=
\Pi_\mp+\Gamma_2+\Gamma_2^f.
\eea

\bea
&\epsilon\times\bar{\Sigma}_1
=
\bar{\Sigma}_1+\bar{\Sigma}_2+\bar{\Sigma}_2^f,\\
&\epsilon\times\bar{\Sigma}_2
=
\bar{\Sigma}_1+\bar{\Sigma}_1^f
+3\bar{\Sigma}_2+2\bar{\Sigma}_2^f,\\
&\tau_\pm\times\bar{\Sigma}_1
=
\bar{\Sigma}_2,\\
&\tau_\pm\times\bar{\Sigma}_2
=
\bar{\Sigma}_1+\bar{\Sigma}_2+\bar{\Sigma}_2^f.
\eea

\bea
&\Sigma_1\times\Sigma_1
=
\Gamma_1^f+\Gamma_2,\\
&\Sigma_1\times\Sigma_2
=
\Gamma_1+\Gamma_2+\Pi_++\Pi_-,\\
&\Sigma_2\times\Sigma_2
=
\Gamma_1+2\Gamma_1^f
+2\Gamma_2+\Gamma_2^f
+2\Pi_+ +2\Pi_-.
\eea

\bea
&\Sigma_1\times\Gamma_1
=
\bar{\Sigma}_1+\bar{\Sigma}_2^f,\\
&\Sigma_1\times\Gamma_2
=
\bar{\Sigma}_1+\bar{\Sigma}_2,\\
&\Sigma_2\times\Gamma_1
=
\bar{\Sigma}_1^f+2\bar{\Sigma}_2+\bar{\Sigma}_2^f,\\
&\Sigma_2\times\Gamma_2
=
\bar{\Sigma}_1+2\bar{\Sigma}_2+\bar{\Sigma}_2^f,\\
&\Sigma_1\times\Pi_\pm
=
\bar{\Sigma}_2+\bar{\Sigma}_2^f,\\
&\Sigma_2\times\Pi_\pm
=
\bar{\Sigma}_1+\bar{\Sigma}_1^f
+2\bar{\Sigma}_2+2\bar{\Sigma}_2^f.
\eea

\bea
&\Sigma_1\times\bar{\Sigma}_1
=
1+\epsilon,\\
&\Sigma_1\times\bar{\Sigma}_2
=
\epsilon+\epsilon^f+\tau_++\tau_-,\\
&\Sigma_2\times\bar{\Sigma}_1
=
\epsilon+\epsilon^f+\tau_++\tau_-,\\
&\Sigma_2\times\bar{\Sigma}_2
=
1+3\epsilon+2\epsilon^f
+\tau_++\tau_-+\tau_+^f+\tau_-^f.
\eea

\bea
&\Gamma_1\times\Gamma_1
=
f+\epsilon^f+\tau_++\tau_-,\\
&\Gamma_2\times\Gamma_2
=
1+\epsilon+\tau_++\tau_-,\\
&\Gamma_1\times\Gamma_2
=
\epsilon+\epsilon^f,\\
&\Gamma_1\times\Pi_\pm
=
\epsilon+\epsilon^f+\tau_\pm+\tau_\pm^f,\\
&\Gamma_2\times\Pi_\pm
=
\epsilon+\epsilon^f+\tau_\mp+\tau_\mp^f,\\
&\Pi_\pm\times\Pi_\pm
=
1+f+\epsilon+\epsilon^f
+\tau_++\tau_-+\tau_+^f+\tau_-^f,\\
&\Pi_+\times\Pi_-
=
2\epsilon+2\epsilon^f.
\eea

\bea
&\Gamma_1\times\bar{\Sigma}_1
=
\Sigma_1^f+\Sigma_2,\\
&\Gamma_1\times\bar{\Sigma}_2
=
\Sigma_1+\Sigma_2+2\Sigma_2^f,\\
&\Gamma_2\times\bar{\Sigma}_1
=
\Sigma_1+\Sigma_2,\\
&\Gamma_2\times\bar{\Sigma}_2
=
\Sigma_1+2\Sigma_2+\Sigma_2^f,\\
&\Pi_\pm\times\bar{\Sigma}_1
=
\Sigma_2+\Sigma_2^f,\\
&\Pi_\pm\times\bar{\Sigma}_2
=
\Sigma_1+\Sigma_1^f
+2\Sigma_2+2\Sigma_2^f.
\eea

\bea
&\bar{\Sigma}_1\times\bar{\Sigma}_1
=
\Gamma_1+\Gamma_2,\\
&\bar{\Sigma}_1\times\bar{\Sigma}_2
=
\Gamma_1^f+\Gamma_2+\Pi_++\Pi_-,\\
&\bar{\Sigma}_2\times\bar{\Sigma}_2
=
2\Gamma_1+\Gamma_1^f
+2\Gamma_2+\Gamma_2^f
+2\Pi_+ +2\Pi_-.
\eea

\section{Fusion rules of charge-$4e$ primitive SC$^\sharp$ from $\mathrm{Spin}(10)_4$}\label{app:fusion2}

Anyons in $\mathrm{Spin}(10)_4$ CS theory are labeled by their $\mathrm{Spin}(10)$ weight $(\lambda_1,\lambda_2,\lambda_3,\lambda_4,\lambda_5)$, with nonnegative integers $\lambda_i$ satisfying $\lambda_1+2\lambda_2+2\lambda_3+\lambda_4+\lambda_5\le 4$. The fermionic simple current is $\Lambda=(0,0,0,0,4)$ with topological spin $5/2$. The fusion of an anyon in $\mathrm{Spin}(10)_4$ with $\Lambda$ is $\Lambda\times (\lambda_1,\lambda_2,\lambda_3,\lambda_4,\lambda_5)=(\lambda_5,\lambda_3,\lambda_2,\lambda_1,\lambda_0)$, where $\lambda_0=4-(\lambda_1+2\lambda_2+2\lambda_3+\lambda_4+\lambda_5)$. The monodromy between an anyon and $\Lambda$ is $M=i^{2\lambda_1+2\lambda_3+3\lambda_4+\lambda_5}$. Consequently, condensing $[(0,0,0,0,4),\mathbf{1}]$ in $\mathrm{Spin}(10)_4\boxtimes\mathrm{sRep}(U(1))$ results in the following sectors (physical electron $f$ omitted). Similar to the $SU(4)_4$ case, orbits with length-2 split into two anyons. Orbits with length-1 also split into two objects that are invariant upon fusing $f$. They appear in the $h/(2e)$ sector and are labeled by $\Pi_\pm$ and $\Xi_\pm$.

\onecolumngrid

\begin{itemize}
    \item The dynamical anyon sector:
    \bea
    & 1 \sim(0,0,0,0,0)\sim(0,0,0,0,4)\sim(4,0,0,0,0)\sim(0,0,0,4,0),\\
    & \epsilon_1 \sim(0,0,1,0,2)\sim(2,1,0,0,0)\sim(0,0,1,2,0)\sim(0,1,0,0,0),\\
    & \epsilon_2 \sim(0,0,0,1,1)\sim(1,0,0,0,2)\sim(2,0,0,1,1)\sim(1,0,0,2,0),\\
    & \tau_\pm \sim (0,0,0,2,2)\sim(2,0,0,0,0),\\
    & \sigma_\pm \sim (0,0,2,0,0)\sim(0,2,0,0,0),\\
    & \mu_\pm \sim (0,1,0,1,1)\sim(1,0,1,0,0),
    \eea
    \item The $h/(4e)$ vortex sector:
    \bea
    & \Sigma_1 \sim(0,0,0,0,1)\sim(1,0,0,0,3)\sim(3,0,0,1,0)\sim(0,0,0,3,0),\\
    & \Sigma_2 \sim(0,0,0,1,2)\sim(2,0,0,0,1)\sim(1,0,0,2,1)\sim(1,0,0,1,0),\\
    & \Sigma_3 \sim(0,0,1,1,0)\sim(0,1,0,0,1)\sim(1,0,1,0,1)\sim(1,1,0,1,0).
    \eea
    \item The $h/(2e)$ vortex sector:
    \bea
    & \Gamma_1 \sim(0,0,0,1,3)\sim(3,0,0,0,0)\sim(0,0,0,3,1)\sim(1,0,0,0,0),\\
    & \Gamma_2 \sim(0,0,0,0,2)\sim(2,0,0,0,2)\sim(2,0,0,2,0)\sim(0,0,0,2,0),\\
    & \Gamma_3 \sim(0,0,1,0,0)\sim(0,1,0,0,2)\sim(2,0,1,0,0)\sim(0,1,0,2,0),\\
    & \Omega_\pm \sim(0,0,1,1,1)\sim(1,1,0,0,0),\\
    & \Pi_\pm = f\times \Pi_\pm \sim(0,1,1,0,0),\\
    & \Xi_\pm = f\times \Xi_\pm \sim(1,0,0,1,1).
    \eea
    \item The $3h/(4e)$ vortex sector:
    \bea
    & \bar{\Sigma}_1 \sim(0,0,0,0,3)\sim(3,0,0,0,1)\sim(1,0,0,3,0)\sim(0,0,0,1,0),\\
    & \bar{\Sigma}_2 \sim(0,0,0,2,1)\sim(1,0,0,0,1)\sim(1,0,0,1,2)\sim(2,0,0,1,0),\\
    & \bar{\Sigma}_3 \sim(0,0,1,0,1)\sim(1,1,0,0,1)\sim(1,0,1,1,0)\sim(0,1,0,1,0).
    \eea
\end{itemize}


The fusion rules are as follows. In parallel to $SU(4)_4$, superscript $f$ denotes fusion with physical electron $f$. 
\bea
&\epsilon_1\times \epsilon_1
=
1+\epsilon_1^f+\epsilon_2+\tau_+^f+\tau_-^f+\sigma_+^f+\sigma_-^f+\mu_+^f+\mu_-^f,\\
&\epsilon_1\times \epsilon_2
=
\epsilon_1+2\epsilon_2+2\epsilon_2^f+\mu_++\mu_+^f+\mu_-+\mu_-^f,\\
&\epsilon_1\times \tau_+
=
\epsilon_1^f+\tau_-^f+\mu_+^f,\\
&\epsilon_1\times \tau_-
=
\epsilon_1^f+\tau_+^f+\mu_-^f,\\
&\epsilon_1\times \sigma_+
=
\epsilon_1^f+\mu_++\mu_+^f,\\
&\epsilon_1\times \sigma_-
=
\epsilon_1^f+\mu_-+\mu_-^f,\\
&\epsilon_1\times \mu_+
=
\epsilon_1^f+\epsilon_2+\epsilon_2^f+\tau_+^f+\sigma_++\sigma_+^f+\mu_++\mu_+^f+\mu_-^f,\\
&\epsilon_1\times \mu_-
=
\epsilon_1^f+\epsilon_2+\epsilon_2^f+\tau_-^f+\sigma_-+\sigma_-^f+\mu_+^f+\mu_-+\mu_-^f,\\
&\epsilon_2\times \epsilon_2
=
1+2\epsilon_1+2\epsilon_1^f+2\epsilon_2+2\epsilon_2^f+\tau_++\tau_+^f+\tau_-+\tau_-^f+\sigma_++\sigma_+^f+\sigma_-+\sigma_-^f+2\mu_++2\mu_+^f+2\mu_-+2\mu_-^f,\\
&\epsilon_2\times \tau_+
=
\epsilon_2+\epsilon_2^f+\mu_-,\\
&\epsilon_2\times \tau_-
=
\epsilon_2+\epsilon_2^f+\mu_+,\\
&\epsilon_2\times \sigma_+
=
\epsilon_2+\epsilon_2^f+\sigma_-+\mu_-+\mu_-^f,\\
&\epsilon_2\times \sigma_-
=
\epsilon_2+\epsilon_2^f+\sigma_++\mu_++\mu_+^f,\\
&\epsilon_2\times \mu_+
=
\epsilon_1+\epsilon_1^f+2\epsilon_2+2\epsilon_2^f+\tau_-+\sigma_-+\sigma_-^f+\mu_++\mu_+^f+2\mu_-+2\mu_-^f,\\
&\epsilon_2\times \mu_-
=
\epsilon_1+\epsilon_1^f+2\epsilon_2+2\epsilon_2^f+\tau_++\sigma_++\sigma_+^f+2\mu_++2\mu_+^f+\mu_-+\mu_-^f,\\
\eea

\bea
&\tau_+\times \tau_+
=
1+\tau_+^f+\sigma_+^f,\\
&\tau_+\times \tau_-
=
\epsilon_1^f,\\
&\tau_+\times \sigma_+
=
\tau_+^f+\sigma_++\sigma_+^f,\\
&\tau_+\times \sigma_-
=
\mu_-^f,\\
&\tau_+\times \mu_+
=
\epsilon_1^f+\mu_++\mu_+^f,\\
&\tau_+\times \mu_-
=
\epsilon_2+\sigma_-^f+\mu_-^f,\\
&\tau_-\times \tau_-
=
1+\tau_-^f+\sigma_-^f,\\
&\tau_-\times \sigma_+
=
\mu_+^f,\\
&\tau_-\times \sigma_-
=
\tau_-^f+\sigma_-+\sigma_-^f,\\
&\tau_-\times \mu_+
=
\epsilon_2+\sigma_+^f+\mu_+^f,\\
&\tau_-\times \mu_-
=
\epsilon_1^f+\mu_-+\mu_-^f,\\
\eea

\bea
&\sigma_+\times \sigma_+
=
1+\tau_++\tau_+^f+\sigma_++\sigma_+^f,\\
&\sigma_+\times \sigma_-
=
\epsilon_2,\\
&\sigma_+\times \mu_+
=
\epsilon_1+\epsilon_1^f+\tau_-^f+\mu_++\mu_+^f,\\
&\sigma_+\times \mu_-
=
\epsilon_2+\epsilon_2^f+\mu_-^f,\\
&\sigma_-\times \sigma_-
=
1+\tau_-+\tau_-^f+\sigma_-+\sigma_-^f,\\
&\sigma_-\times \mu_+
=
\epsilon_2+\epsilon_2^f+\mu_+^f,\\
&\sigma_-\times \mu_-
=
\epsilon_1+\epsilon_1^f+\tau_+^f+\mu_-+\mu_-^f,\\
\eea

\bea
&\mu_+\times \mu_+
=
1+\epsilon_1+\epsilon_1^f+\epsilon_2+\epsilon_2^f+\tau_++\tau_+^f+\tau_-^f+\sigma_++\sigma_+^f+\sigma_-^f+\mu_++\mu_+^f+\mu_-+\mu_-^f,\\
&\mu_+\times \mu_-
=
\epsilon_1^f+2\epsilon_2+2\epsilon_2^f+\mu_++\mu_+^f+\mu_-+\mu_-^f,\\
&\mu_-\times \mu_-
=
1+\epsilon_1+\epsilon_1^f+\epsilon_2+\epsilon_2^f+\tau_+^f+\tau_-+\tau_-^f+\sigma_+^f+\sigma_-+\sigma_-^f+\mu_++\mu_+^f+\mu_-+\mu_-^f.
\eea

\bea
&\epsilon_1\times \Sigma_1
=
\Sigma_1^f+\Sigma_2+\Sigma_3,\\
&\epsilon_1\times \Sigma_2
=
\Sigma_1+3\Sigma_2^f+\Sigma_3+2\Sigma_3^f,\\
&\epsilon_1\times \Sigma_3
=
\Sigma_1+\Sigma_2+2\Sigma_2^f+2\Sigma_3+3\Sigma_3^f,\\
&\epsilon_2\times \Sigma_1
=
\Sigma_1+\Sigma_2+\Sigma_2^f+\Sigma_3+\Sigma_3^f,\\
&\epsilon_2\times \Sigma_2
=
\Sigma_1+\Sigma_1^f+3\Sigma_2+2\Sigma_2^f+3\Sigma_3+3\Sigma_3^f,\\
&\epsilon_2\times \Sigma_3
=
\Sigma_1+\Sigma_1^f+3\Sigma_2+3\Sigma_2^f+5\Sigma_3+4\Sigma_3^f,\\
\eea

\bea
&\tau_+\times \Sigma_1
=
\Sigma_2,\\
&\tau_+\times \Sigma_2
=
\Sigma_1+\Sigma_2^f+\Sigma_3^f,\\
&\tau_+\times \Sigma_3
=
\Sigma_2^f+\Sigma_3+\Sigma_3^f,\\
&\tau_-\times \Sigma_1
=
\Sigma_2,\\
&\tau_-\times \Sigma_2
=
\Sigma_1+\Sigma_2^f+\Sigma_3^f,\\
&\tau_-\times \Sigma_3
=
\Sigma_2^f+\Sigma_3+\Sigma_3^f,\\
\eea

\bea
&\sigma_+\times \Sigma_1
=
\Sigma_3,\\
&\sigma_+\times \Sigma_2
=
\Sigma_2^f+\Sigma_3+\Sigma_3^f,\\
&\sigma_+\times \Sigma_3
=
\Sigma_1+\Sigma_2+\Sigma_2^f+\Sigma_3+\Sigma_3^f,\\
&\sigma_-\times \Sigma_1
=
\Sigma_3,\\
&\sigma_-\times \Sigma_2
=
\Sigma_2^f+\Sigma_3+\Sigma_3^f,\\
&\sigma_-\times \Sigma_3
=
\Sigma_1+\Sigma_2+\Sigma_2^f+\Sigma_3+\Sigma_3^f,\\
\eea

\bea
&\mu_+\times \Sigma_1
=
\Sigma_2+\Sigma_3+\Sigma_3^f,\\
&\mu_+\times \Sigma_2
=
\Sigma_1+\Sigma_2+2\Sigma_2^f+2\Sigma_3+3\Sigma_3^f,\\
&\mu_+\times \Sigma_3
=
\Sigma_1+\Sigma_1^f+2\Sigma_2+3\Sigma_2^f+3\Sigma_3+3\Sigma_3^f,\\
&\mu_-\times \Sigma_1
=
\Sigma_2+\Sigma_3+\Sigma_3^f,\\
&\mu_-\times \Sigma_2
=
\Sigma_1+\Sigma_2+2\Sigma_2^f+2\Sigma_3+3\Sigma_3^f,\\
&\mu_-\times \Sigma_3
=
\Sigma_1+\Sigma_1^f+2\Sigma_2+3\Sigma_2^f+3\Sigma_3+3\Sigma_3^f.
\eea

\bea
&\epsilon_1\times \Gamma_1
=
\Gamma_1^f+\Gamma_3+\Omega_+^f+\Omega_-^f,\\
&\epsilon_1\times \Gamma_2
=
\Gamma_2^f+\Gamma_3+\Gamma_3^f+\Xi_++\Xi_-,\\
&\epsilon_1\times \Gamma_3
=
\Gamma_1+\Gamma_2+\Gamma_2^f+\Gamma_3^f+\Omega_++\Omega_-+\Pi_++\Pi_-+\Xi_++\Xi_-,\\
&\epsilon_1\times \Omega_+
=
\Gamma_1^f+\Gamma_3+\Omega_+^f+\Omega_-^f+\Pi_++\Xi_+,\\
&\epsilon_1\times \Omega_-
=
\Gamma_1^f+\Gamma_3+\Omega_+^f+\Omega_-^f+\Pi_-+\Xi_-,\\
&\epsilon_1\times \Pi_+
=
\Gamma_3+\Gamma_3^f+\Omega_++\Omega_+^f+\Pi_++\Xi_+,\\
&\epsilon_1\times \Pi_-
=
\Gamma_3+\Gamma_3^f+\Omega_-+\Omega_-^f+\Pi_-+\Xi_-,\\
&\epsilon_1\times \Xi_+
=
\Gamma_2+\Gamma_2^f+\Gamma_3+\Gamma_3^f+\Omega_++\Omega_+^f+\Pi_++\Xi_++2\Xi_-,\\
&\epsilon_1\times \Xi_-
=
\Gamma_2+\Gamma_2^f+\Gamma_3+\Gamma_3^f+\Omega_-+\Omega_-^f+\Pi_-+2\Xi_++\Xi_-,\\
\eea

\bea
&\epsilon_2\times \Gamma_1
=
\Gamma_2+\Gamma_2^f+\Gamma_3^f+\Xi_++\Xi_-,\\
&\epsilon_2\times \Gamma_2
=
\Gamma_1+\Gamma_1^f+\Gamma_2+\Gamma_3+\Gamma_3^f+\Omega_++\Omega_+^f+\Omega_-+\Omega_-^f+\Pi_++\Pi_-+\Xi_++\Xi_-,\\
&\epsilon_2\times \Gamma_3
=
\Gamma_1^f+\Gamma_2+\Gamma_2^f+2\Gamma_3+2\Gamma_3^f+\Omega_++\Omega_+^f+\Omega_-+\Omega_-^f+\Pi_++\Pi_-+2\Xi_++2\Xi_-,\\
&\epsilon_2\times \Omega_+
=
\Gamma_2+\Gamma_2^f+\Gamma_3+\Gamma_3^f+\Omega_-+\Pi_-+\Xi_++2\Xi_-,\\
&\epsilon_2\times \Omega_-
=
\Gamma_2+\Gamma_2^f+\Gamma_3+\Gamma_3^f+\Omega_++\Pi_++2\Xi_++\Xi_-,\\
&\epsilon_2\times \Pi_+
=
\Gamma_2+\Gamma_2^f+\Gamma_3+\Gamma_3^f+\Omega_-+\Omega_-^f+2\Pi_-+\Xi_++2\Xi_-,\\
&\epsilon_2\times \Pi_-
=
\Gamma_2+\Gamma_2^f+\Gamma_3+\Gamma_3^f+\Omega_++\Omega_+^f+2\Pi_++2\Xi_++\Xi_-,\\
&\epsilon_2\times \Xi_+
=
\Gamma_1+\Gamma_1^f+\Gamma_2+\Gamma_2^f+2\Gamma_3+2\Gamma_3^f+\Omega_++\Omega_+^f+2\Omega_-+2\Omega_-^f+\Pi_++2\Pi_-+2\Xi_++2\Xi_-,\\
&\epsilon_2\times \Xi_-
=
\Gamma_1+\Gamma_1^f+\Gamma_2+\Gamma_2^f+2\Gamma_3+2\Gamma_3^f+2\Omega_++2\Omega_+^f+\Omega_-+\Omega_-^f+2\Pi_++\Pi_-+2\Xi_++2\Xi_-,\\
\eea

\bea
&\tau_+\times \Gamma_1
=
\Gamma_1^f+\Omega_+^f,\\
&\tau_+\times \Gamma_2
=
\Gamma_2+\Xi_-,\\
&\tau_+\times \Gamma_3
=
\Gamma_3^f+\Omega_-+\Xi_+,\\
&\tau_+\times \Omega_+
=
\Gamma_1^f+\Omega_+^f+\Pi_+,\\
&\tau_+\times \Omega_-
=
\Gamma_3+\Omega_-^f,\\
&\tau_+\times \Pi_+
=
\Omega_++\Omega_+^f+\Pi_+,\\
&\tau_+\times \Pi_-
=
\Pi_-+\Xi_-,\\
&\tau_+\times \Xi_+
=
\Gamma_3+\Gamma_3^f+\Xi_+,\\
&\tau_+\times \Xi_-
=
\Gamma_2+\Gamma_2^f+\Pi_-+\Xi_-,\\
\eea

\bea
&\tau_-\times \Gamma_1
=
\Gamma_1^f+\Omega_-^f,\\
&\tau_-\times \Gamma_2
=
\Gamma_2+\Xi_+,\\
&\tau_-\times \Gamma_3
=
\Gamma_3^f+\Omega_++\Xi_-,\\
&\tau_-\times \Omega_+
=
\Gamma_3+\Omega_+^f,\\
&\tau_-\times \Omega_-
=
\Gamma_1^f+\Omega_-^f+\Pi_-,\\
&\tau_-\times \Pi_+
=
\Pi_++\Xi_+,\\
&\tau_-\times \Pi_-
=
\Omega_-+\Omega_-^f+\Pi_-,\\
&\tau_-\times \Xi_+
=
\Gamma_2+\Gamma_2^f+\Pi_++\Xi_+,\\
&\tau_-\times \Xi_-
=
\Gamma_3+\Gamma_3^f+\Xi_-,\\
\eea

\bea
&\sigma_+\times \Gamma_1
=
\Omega_+^f+\Pi_+,\\
&\sigma_+\times \Gamma_2
=
\Gamma_2^f+\Pi_-+\Xi_-,\\
&\sigma_+\times \Gamma_3
=
\Gamma_3+\Gamma_3^f+\Omega_-+\Xi_+,\\
&\sigma_+\times \Omega_+
=
\Gamma_1^f+\Omega_++\Omega_+^f+\Pi_+,\\
&\sigma_+\times \Omega_-
=
\Gamma_3+\Xi_+,\\
&\sigma_+\times \Pi_+
=
\Gamma_1+\Gamma_1^f+\Omega_++\Omega_+^f+\Pi_+,\\
&\sigma_+\times \Pi_-
=
\Gamma_2+\Gamma_2^f+\Xi_-,\\
&\sigma_+\times \Xi_+
=
\Gamma_3+\Gamma_3^f+\Omega_-+\Omega_-^f+\Xi_+,\\
&\sigma_+\times \Xi_-
=
\Gamma_2+\Gamma_2^f+\Pi_-+2\Xi_-,\\
\eea

\bea
&\sigma_-\times \Gamma_1
=
\Omega_-^f+\Pi_-,\\
&\sigma_-\times \Gamma_2
=
\Gamma_2^f+\Pi_++\Xi_+,\\
&\sigma_-\times \Gamma_3
=
\Gamma_3+\Gamma_3^f+\Omega_++\Xi_-,\\
&\sigma_-\times \Omega_+
=
\Gamma_3+\Xi_-,\\
&\sigma_-\times \Omega_-
=
\Gamma_1^f+\Omega_-+\Omega_-^f+\Pi_-,\\
&\sigma_-\times \Pi_+
=
\Gamma_2+\Gamma_2^f+\Xi_+,\\
&\sigma_-\times \Pi_-
=
\Gamma_1+\Gamma_1^f+\Omega_-+\Omega_-^f+\Pi_-,\\
&\sigma_-\times \Xi_+
=
\Gamma_2+\Gamma_2^f+\Pi_++2\Xi_+,\\
&\sigma_-\times \Xi_-
=
\Gamma_3+\Gamma_3^f+\Omega_++\Omega_+^f+\Xi_-,\\
\eea

\bea
&\mu_+\times \Gamma_1
=
\Gamma_3+\Omega_+^f+\Pi_++\Xi_+,\\
&\mu_+\times \Gamma_2
=
\Gamma_2+\Gamma_3+\Gamma_3^f+\Omega_-+\Omega_-^f+\Pi_-+\Xi_++\Xi_-,\\
&\mu_+\times \Gamma_3
=
\Gamma_1+\Gamma_2+\Gamma_2^f+\Gamma_3+\Gamma_3^f+\Omega_++\Omega_+^f+\Omega_-+\Pi_++\Pi_-+\Xi_++2\Xi_-,\\
&\mu_+\times \Omega_+
=
\Gamma_1^f+\Gamma_3+\Gamma_3^f+\Omega_++\Omega_+^f+\Omega_-^f+\Pi_++\Xi_+,\\
&\mu_+\times \Omega_-
=
\Gamma_2+\Gamma_2^f+\Gamma_3+\Omega_+^f+\Pi_++\Xi_++\Xi_-,\\
&\mu_+\times \Pi_+
=
\Gamma_1+\Gamma_1^f+\Gamma_3+\Gamma_3^f+\Omega_++\Omega_+^f+\Omega_-+\Omega_-^f+\Pi_++\Xi_+,\\
&\mu_+\times \Pi_-
=
\Gamma_2+\Gamma_2^f+\Gamma_3+\Gamma_3^f+2\Xi_++\Xi_-,\\
&\mu_+\times \Xi_+
=
\Gamma_1+\Gamma_1^f+\Gamma_2+\Gamma_2^f+\Gamma_3+\Gamma_3^f+\Omega_++\Omega_+^f+\Omega_-+\Omega_-^f+\Pi_++2\Pi_-+\Xi_++2\Xi_-,\\
&\mu_+\times \Xi_-
=
\Gamma_2+\Gamma_2^f+2\Gamma_3+2\Gamma_3^f+\Omega_-+\Omega_-^f+\Pi_-+2\Xi_++2\Xi_-,\\
\eea

\bea
&\mu_-\times \Gamma_1
=
\Gamma_3+\Omega_-^f+\Pi_-+\Xi_-,\\
&\mu_-\times \Gamma_2
=
\Gamma_2+\Gamma_3+\Gamma_3^f+\Omega_++\Omega_+^f+\Pi_++\Xi_++\Xi_-,\\
&\mu_-\times \Gamma_3
=
\Gamma_1+\Gamma_2+\Gamma_2^f+\Gamma_3+\Gamma_3^f+\Omega_++\Omega_-+\Omega_-^f+\Pi_++\Pi_-+2\Xi_++\Xi_-,\\
&\mu_-\times \Omega_+
=
\Gamma_2+\Gamma_2^f+\Gamma_3+\Omega_-^f+\Pi_-+\Xi_++\Xi_-,\\
&\mu_-\times \Omega_-
=
\Gamma_1^f+\Gamma_3+\Gamma_3^f+\Omega_+^f+\Omega_-+\Omega_-^f+\Pi_-+\Xi_-,\\
&\mu_-\times \Pi_+
=
\Gamma_2+\Gamma_2^f+\Gamma_3+\Gamma_3^f+\Xi_++2\Xi_-,\\
&\mu_-\times \Pi_-
=
\Gamma_1+\Gamma_1^f+\Gamma_3+\Gamma_3^f+\Omega_++\Omega_+^f+\Omega_-+\Omega_-^f+\Pi_-+\Xi_-,\\
&\mu_-\times \Xi_+
=
\Gamma_2+\Gamma_2^f+2\Gamma_3+2\Gamma_3^f+\Omega_++\Omega_+^f+\Pi_++2\Xi_++2\Xi_-,\\
&\mu_-\times \Xi_-
=
\Gamma_1+\Gamma_1^f+\Gamma_2+\Gamma_2^f+\Gamma_3+\Gamma_3^f+\Omega_++\Omega_+^f+\Omega_-+\Omega_-^f+2\Pi_++\Pi_-+2\Xi_++\Xi_-.
\eea

\bea
&\epsilon_1\times \bar{\Sigma}_1
=
\bar{\Sigma}_1^f+\bar{\Sigma}_2^f+\bar{\Sigma}_3^f,\\
&\epsilon_1\times \bar{\Sigma}_2
=
\bar{\Sigma}_1^f+3\bar{\Sigma}_2^f+\bar{\Sigma}_3+2\bar{\Sigma}_3^f,\\
&\epsilon_1\times \bar{\Sigma}_3
=
\bar{\Sigma}_1^f+\bar{\Sigma}_2+2\bar{\Sigma}_2^f+2\bar{\Sigma}_3+3\bar{\Sigma}_3^f,\\
&\epsilon_2\times \bar{\Sigma}_1
=
\bar{\Sigma}_1+\bar{\Sigma}_2+\bar{\Sigma}_2^f+\bar{\Sigma}_3+\bar{\Sigma}_3^f,\\
&\epsilon_2\times \bar{\Sigma}_2
=
\bar{\Sigma}_1+\bar{\Sigma}_1^f+3\bar{\Sigma}_2+2\bar{\Sigma}_2^f+3\bar{\Sigma}_3+3\bar{\Sigma}_3^f,\\
&\epsilon_2\times \bar{\Sigma}_3
=
\bar{\Sigma}_1+\bar{\Sigma}_1^f+3\bar{\Sigma}_2+3\bar{\Sigma}_2^f+5\bar{\Sigma}_3+4\bar{\Sigma}_3^f,\\
\eea

\bea
&\tau_+\times \bar{\Sigma}_1
=
\bar{\Sigma}_2^f,\\
&\tau_+\times \bar{\Sigma}_2
=
\bar{\Sigma}_1^f+\bar{\Sigma}_2^f+\bar{\Sigma}_3^f,\\
&\tau_+\times \bar{\Sigma}_3
=
\bar{\Sigma}_2^f+\bar{\Sigma}_3+\bar{\Sigma}_3^f,\\
&\tau_-\times \bar{\Sigma}_1
=
\bar{\Sigma}_2^f,\\
&\tau_-\times \bar{\Sigma}_2
=
\bar{\Sigma}_1^f+\bar{\Sigma}_2^f+\bar{\Sigma}_3^f,\\
&\tau_-\times \bar{\Sigma}_3
=
\bar{\Sigma}_2^f+\bar{\Sigma}_3+\bar{\Sigma}_3^f,\\
\eea

\bea
&\sigma_+\times \bar{\Sigma}_1
=
\bar{\Sigma}_3^f,\\
&\sigma_+\times \bar{\Sigma}_2
=
\bar{\Sigma}_2^f+\bar{\Sigma}_3+\bar{\Sigma}_3^f,\\
&\sigma_+\times \bar{\Sigma}_3
=
\bar{\Sigma}_1^f+\bar{\Sigma}_2+\bar{\Sigma}_2^f+\bar{\Sigma}_3+\bar{\Sigma}_3^f,\\
&\sigma_-\times \bar{\Sigma}_1
=
\bar{\Sigma}_3^f,\\
&\sigma_-\times \bar{\Sigma}_2
=
\bar{\Sigma}_2^f+\bar{\Sigma}_3+\bar{\Sigma}_3^f,\\
&\sigma_-\times \bar{\Sigma}_3
=
\bar{\Sigma}_1^f+\bar{\Sigma}_2+\bar{\Sigma}_2^f+\bar{\Sigma}_3+\bar{\Sigma}_3^f,\\
\eea

\bea
&\mu_+\times \bar{\Sigma}_1
=
\bar{\Sigma}_2^f+\bar{\Sigma}_3+\bar{\Sigma}_3^f,\\
&\mu_+\times \bar{\Sigma}_2
=
\bar{\Sigma}_1^f+\bar{\Sigma}_2+2\bar{\Sigma}_2^f+2\bar{\Sigma}_3+3\bar{\Sigma}_3^f,\\
&\mu_+\times \bar{\Sigma}_3
=
\bar{\Sigma}_1+\bar{\Sigma}_1^f+2\bar{\Sigma}_2+3\bar{\Sigma}_2^f+3\bar{\Sigma}_3+3\bar{\Sigma}_3^f,\\
&\mu_-\times \bar{\Sigma}_1
=
\bar{\Sigma}_2^f+\bar{\Sigma}_3+\bar{\Sigma}_3^f,\\
&\mu_-\times \bar{\Sigma}_2
=
\bar{\Sigma}_1^f+\bar{\Sigma}_2+2\bar{\Sigma}_2^f+2\bar{\Sigma}_3+3\bar{\Sigma}_3^f,\\
&\mu_-\times \bar{\Sigma}_3
=
\bar{\Sigma}_1+\bar{\Sigma}_1^f+2\bar{\Sigma}_2+3\bar{\Sigma}_2^f+3\bar{\Sigma}_3+3\bar{\Sigma}_3^f.
\eea

\bea
&\Sigma_1\times \Sigma_1
=
\Gamma_1^f+\Gamma_2+\Gamma_3,\\
&\Sigma_1\times \Sigma_2
=
\Gamma_1+\Gamma_2+\Gamma_3^f+\Omega_++\Omega_-+\Xi_++\Xi_-,\\
&\Sigma_1\times \Sigma_3
=
\Gamma_2^f+\Gamma_3+\Gamma_3^f+\Omega_++\Omega_-+\Pi_++\Pi_-+\Xi_++\Xi_-,\\
&\Sigma_2\times \Sigma_2
=
2\Gamma_1^f+2\Gamma_2+\Gamma_2^f+3\Gamma_3+\Gamma_3^f+2\Omega_+^f+2\Omega_-^f+\Pi_++\Pi_-+2\Xi_++2\Xi_-,\\
&\Sigma_2\times \Sigma_3
=
\Gamma_1^f+\Gamma_2+2\Gamma_2^f+3\Gamma_3+2\Gamma_3^f+\Omega_++2\Omega_+^f+\Omega_-+2\Omega_-^f+2\Pi_++2\Pi_-+3\Xi_++3\Xi_-,\\
&\Sigma_3\times \Sigma_3
=
\Gamma_1+2\Gamma_1^f+3\Gamma_2+2\Gamma_2^f+4\Gamma_3+3\Gamma_3^f+2\Omega_++2\Omega_+^f+2\Omega_-+2\Omega_-^f+2\Pi_++2\Pi_-+4\Xi_++4\Xi_-.
\eea

\bea
&\Sigma_1\times \Gamma_1
=
\bar{\Sigma}_1+\bar{\Sigma}_2,\\
&\Sigma_1\times \Gamma_2
=
\bar{\Sigma}_1+\bar{\Sigma}_2^f+\bar{\Sigma}_3,\\
&\Sigma_1\times \Gamma_3
=
\bar{\Sigma}_1^f+\bar{\Sigma}_2^f+\bar{\Sigma}_3+\bar{\Sigma}_3^f,\\
&\Sigma_1\times \Omega_+
=
\bar{\Sigma}_2+\bar{\Sigma}_3,\\
&\Sigma_1\times \Omega_-
=
\bar{\Sigma}_2+\bar{\Sigma}_3,\\
&\Sigma_1\times \Pi_+
=
\bar{\Sigma}_3+\bar{\Sigma}_3^f,\\
&\Sigma_1\times \Pi_-
=
\bar{\Sigma}_3+\bar{\Sigma}_3^f,\\
&\Sigma_1\times \Xi_+
=
\bar{\Sigma}_2+\bar{\Sigma}_2^f+\bar{\Sigma}_3+\bar{\Sigma}_3^f,\\
&\Sigma_1\times \Xi_-
=
\bar{\Sigma}_2+\bar{\Sigma}_2^f+\bar{\Sigma}_3+\bar{\Sigma}_3^f,\\
\eea

\bea
&\Sigma_2\times \Gamma_1
=
\bar{\Sigma}_1^f+2\bar{\Sigma}_2^f+\bar{\Sigma}_3^f,\\
&\Sigma_2\times \Gamma_2
=
\bar{\Sigma}_1+\bar{\Sigma}_2+2\bar{\Sigma}_2^f+2\bar{\Sigma}_3+\bar{\Sigma}_3^f,\\
&\Sigma_2\times \Gamma_3
=
\bar{\Sigma}_1+3\bar{\Sigma}_2+\bar{\Sigma}_2^f+3\bar{\Sigma}_3+2\bar{\Sigma}_3^f,\\
&\Sigma_2\times \Omega_+
=
\bar{\Sigma}_1^f+2\bar{\Sigma}_2^f+\bar{\Sigma}_3+2\bar{\Sigma}_3^f,\\
&\Sigma_2\times \Omega_-
=
\bar{\Sigma}_1^f+2\bar{\Sigma}_2^f+\bar{\Sigma}_3+2\bar{\Sigma}_3^f,\\
&\Sigma_2\times \Pi_+
=
\bar{\Sigma}_2+\bar{\Sigma}_2^f+2\bar{\Sigma}_3+2\bar{\Sigma}_3^f,\\
&\Sigma_2\times \Pi_-
=
\bar{\Sigma}_2+\bar{\Sigma}_2^f+2\bar{\Sigma}_3+2\bar{\Sigma}_3^f,\\
&\Sigma_2\times \Xi_+
=
\bar{\Sigma}_1+\bar{\Sigma}_1^f+2\bar{\Sigma}_2+2\bar{\Sigma}_2^f+3\bar{\Sigma}_3+3\bar{\Sigma}_3^f,\\
&\Sigma_2\times \Xi_-
=
\bar{\Sigma}_1+\bar{\Sigma}_1^f+2\bar{\Sigma}_2+2\bar{\Sigma}_2^f+3\bar{\Sigma}_3+3\bar{\Sigma}_3^f,\\
\eea

\bea
&\Sigma_3\times \Gamma_1
=
\bar{\Sigma}_2^f+\bar{\Sigma}_3+2\bar{\Sigma}_3^f,\\
&\Sigma_3\times \Gamma_2
=
\bar{\Sigma}_1^f+2\bar{\Sigma}_2+\bar{\Sigma}_2^f+2\bar{\Sigma}_3+3\bar{\Sigma}_3^f,\\
&\Sigma_3\times \Gamma_3
=
\bar{\Sigma}_1+\bar{\Sigma}_1^f+3\bar{\Sigma}_2+2\bar{\Sigma}_2^f+4\bar{\Sigma}_3+3\bar{\Sigma}_3^f,\\
&\Sigma_3\times \Omega_+
=
\bar{\Sigma}_1^f+\bar{\Sigma}_2+2\bar{\Sigma}_2^f+2\bar{\Sigma}_3+2\bar{\Sigma}_3^f,\\
&\Sigma_3\times \Omega_-
=
\bar{\Sigma}_1^f+\bar{\Sigma}_2+2\bar{\Sigma}_2^f+2\bar{\Sigma}_3+2\bar{\Sigma}_3^f,\\
&\Sigma_3\times \Pi_+
=
\bar{\Sigma}_1+\bar{\Sigma}_1^f+2\bar{\Sigma}_2+2\bar{\Sigma}_2^f+2\bar{\Sigma}_3+2\bar{\Sigma}_3^f,\\
&\Sigma_3\times \Pi_-
=
\bar{\Sigma}_1+\bar{\Sigma}_1^f+2\bar{\Sigma}_2+2\bar{\Sigma}_2^f+2\bar{\Sigma}_3+2\bar{\Sigma}_3^f,\\
&\Sigma_3\times \Xi_+
=
\bar{\Sigma}_1+\bar{\Sigma}_1^f+3\bar{\Sigma}_2+3\bar{\Sigma}_2^f+4\bar{\Sigma}_3+4\bar{\Sigma}_3^f,\\
&\Sigma_3\times \Xi_-
=
\bar{\Sigma}_1+\bar{\Sigma}_1^f+3\bar{\Sigma}_2+3\bar{\Sigma}_2^f+4\bar{\Sigma}_3+4\bar{\Sigma}_3^f.
\eea

\bea
&\Sigma_1\times \bar{\Sigma}_1
=
f+\epsilon_1+\epsilon_2^f,\\
&\Sigma_1\times \bar{\Sigma}_2
=
\epsilon_1+\epsilon_2+\epsilon_2^f+\tau_++\tau_-+\mu_++\mu_-,\\
&\Sigma_1\times \bar{\Sigma}_3
=
\epsilon_1+\epsilon_2+\epsilon_2^f+\sigma_++\sigma_-+\mu_++\mu_+^f+\mu_-+\mu_-^f,\\
&\Sigma_2\times \bar{\Sigma}_1
=
\epsilon_1^f+\epsilon_2+\epsilon_2^f+\tau_+^f+\tau_-^f+\mu_+^f+\mu_-^f,\\
&\Sigma_2\times \bar{\Sigma}_2
=
1+3\epsilon_1^f+3\epsilon_2+2\epsilon_2^f+\tau_+^f+\tau_-^f+\sigma_+^f+\sigma_-^f+\mu_++2\mu_+^f+\mu_-+2\mu_-^f,\\
&\Sigma_2\times \bar{\Sigma}_3
=
\epsilon_1+2\epsilon_1^f+3\epsilon_2+3\epsilon_2^f+\tau_+^f+\tau_-^f+\sigma_++\sigma_+^f+\sigma_-+\sigma_-^f+2\mu_++3\mu_+^f+2\mu_-+3\mu_-^f,\\
&\Sigma_3\times \bar{\Sigma}_1
=
\epsilon_1^f+\epsilon_2+\epsilon_2^f+\sigma_+^f+\sigma_-^f+\mu_++\mu_+^f+\mu_-+\mu_-^f,\\
&\Sigma_3\times \bar{\Sigma}_2
=
\epsilon_1+2\epsilon_1^f+3\epsilon_2+3\epsilon_2^f+\tau_+^f+\tau_-^f+\sigma_++\sigma_+^f+\sigma_-+\sigma_-^f+2\mu_++3\mu_+^f+2\mu_-+3\mu_-^f,\\
&\Sigma_3\times \bar{\Sigma}_3
=
1+2\epsilon_1+3\epsilon_1^f+5\epsilon_2+4\epsilon_2^f+\tau_++\tau_+^f+\tau_-+\tau_-^f+\sigma_++\sigma_+^f+\sigma_-+\sigma_-^f+3\mu_++3\mu_+^f+3\mu_-+3\mu_-^f.
\eea

\bea
&\Gamma_1\times \Gamma_1
=
1+\epsilon_1^f+\tau_+^f+\tau_-^f,\\
&\Gamma_1\times \Gamma_2
=
\epsilon_2+\epsilon_2^f,\\
&\Gamma_1\times \Gamma_3
=
\epsilon_1+\epsilon_2^f+\mu_++\mu_-,\\
&\Gamma_1\times \Omega_+
=
\epsilon_1^f+\tau_+^f+\sigma_+^f+\mu_+^f,\\
&\Gamma_1\times \Omega_-
=
\epsilon_1^f+\tau_-^f+\sigma_-^f+\mu_-^f,\\
&\Gamma_1\times \Pi_+
=
\sigma_++\sigma_+^f+\mu_++\mu_+^f,\\
&\Gamma_1\times \Pi_-
=
\sigma_-+\sigma_-^f+\mu_-+\mu_-^f,\\
&\Gamma_1\times \Xi_+
=
\epsilon_2+\epsilon_2^f+\mu_++\mu_+^f,\\
&\Gamma_1\times \Xi_-
=
\epsilon_2+\epsilon_2^f+\mu_-+\mu_-^f,\\
\eea

\bea
&\Gamma_2\times \Gamma_2
=
f+\epsilon_1+\epsilon_2^f+\tau_+^f+\tau_-^f+\sigma_++\sigma_-+\mu_+^f+\mu_-^f,\\
&\Gamma_2\times \Gamma_3
=
\epsilon_1+\epsilon_1^f+\epsilon_2+\epsilon_2^f+\mu_++\mu_+^f+\mu_-+\mu_-^f,\\
&\Gamma_2\times \Omega_+
=
\epsilon_2+\epsilon_2^f+\mu_-+\mu_-^f,\\
&\Gamma_2\times \Omega_-
=
\epsilon_2+\epsilon_2^f+\mu_++\mu_+^f,\\
&\Gamma_2\times \Pi_+
=
\epsilon_2+\epsilon_2^f+\sigma_-+\sigma_-^f+\mu_-+\mu_-^f,\\
&\Gamma_2\times \Pi_-
=
\epsilon_2+\epsilon_2^f+\sigma_++\sigma_+^f+\mu_++\mu_+^f,\\
&\Gamma_2\times \Xi_+
=
\epsilon_1+\epsilon_1^f+\epsilon_2+\epsilon_2^f+\tau_-+\tau_-^f+\sigma_-+\sigma_-^f+\mu_++\mu_+^f+\mu_-+\mu_-^f,\\
&\Gamma_2\times \Xi_-
=
\epsilon_1+\epsilon_1^f+\epsilon_2+\epsilon_2^f+\tau_++\tau_+^f+\sigma_++\sigma_+^f+\mu_++\mu_+^f+\mu_-+\mu_-^f,\\
\eea

\bea
&\Gamma_3\times \Gamma_3
=
1+\epsilon_1^f+2\epsilon_2+2\epsilon_2^f+\tau_+^f+\tau_-^f+\sigma_++\sigma_+^f+\sigma_-+\sigma_-^f+\mu_++\mu_+^f+\mu_-+\mu_-^f,\\
&\Gamma_3\times \Omega_+
=
\epsilon_1+\epsilon_2+\epsilon_2^f+\tau_-+\sigma_-+\mu_++\mu_+^f+\mu_-,\\
&\Gamma_3\times \Omega_-
=
\epsilon_1+\epsilon_2+\epsilon_2^f+\tau_++\sigma_++\mu_++\mu_-+\mu_-^f,\\
&\Gamma_3\times \Pi_+
=
\epsilon_1+\epsilon_1^f+\epsilon_2+\epsilon_2^f+\mu_++\mu_+^f+\mu_-+\mu_-^f,\\
&\Gamma_3\times \Pi_-
=
\epsilon_1+\epsilon_1^f+\epsilon_2+\epsilon_2^f+\mu_++\mu_+^f+\mu_-+\mu_-^f,\\
&\Gamma_3\times \Xi_+
=
\epsilon_1+\epsilon_1^f+2\epsilon_2+2\epsilon_2^f+\tau_++\tau_+^f+\sigma_++\sigma_+^f+\mu_++\mu_+^f+2\mu_-+2\mu_-^f,\\
&\Gamma_3\times \Xi_-
=
\epsilon_1+\epsilon_1^f+2\epsilon_2+2\epsilon_2^f+\tau_-+\tau_-^f+\sigma_-+\sigma_-^f+2\mu_++2\mu_+^f+\mu_-+\mu_-^f,\\
\eea

\bea
&\Omega_+\times \Omega_+
=
1+\epsilon_1^f+\tau_+^f+\tau_-^f+\sigma_++\sigma_+^f+\mu_++\mu_+^f,\\
&\Omega_+\times \Omega_-
=
\epsilon_1^f+\epsilon_2+\mu_+^f+\mu_-^f,\\
&\Omega_+\times \Pi_+
=
\epsilon_1+\epsilon_1^f+\tau_++\tau_+^f+\sigma_++\sigma_+^f+\mu_++\mu_+^f,\\
&\Omega_+\times \Pi_-
=
\epsilon_2+\epsilon_2^f+\mu_-+\mu_-^f,\\
&\Omega_+\times \Xi_+
=
\epsilon_1+\epsilon_1^f+\epsilon_2+\epsilon_2^f+\mu_++\mu_+^f+\mu_-+\mu_-^f,\\
&\Omega_+\times \Xi_-
=
2\epsilon_2+2\epsilon_2^f+\sigma_-+\sigma_-^f+\mu_-+\mu_-^f,\\
&\Omega_-\times \Omega_-
=
1+\epsilon_1^f+\tau_+^f+\tau_-^f+\sigma_-+\sigma_-^f+\mu_-+\mu_-^f,\\
&\Omega_-\times \Pi_+
=
\epsilon_2+\epsilon_2^f+\mu_++\mu_+^f,\\
&\Omega_-\times \Pi_-
=
\epsilon_1+\epsilon_1^f+\tau_-+\tau_-^f+\sigma_-+\sigma_-^f+\mu_-+\mu_-^f,\\
&\Omega_-\times \Xi_+
=
2\epsilon_2+2\epsilon_2^f+\sigma_++\sigma_+^f+\mu_++\mu_+^f,\\
&\Omega_-\times \Xi_-
=
\epsilon_1+\epsilon_1^f+\epsilon_2+\epsilon_2^f+\mu_++\mu_+^f+\mu_-+\mu_-^f,\\
\eea

\bea
&\Pi_+\times \Pi_+
=
1+f+\epsilon_1+\epsilon_1^f+\tau_++\tau_+^f+\tau_-+\tau_-^f+\sigma_++\sigma_+^f+\mu_++\mu_+^f,\\
&\Pi_+\times \Pi_-
=
2\epsilon_2+2\epsilon_2^f,\\
&\Pi_+\times \Xi_+
=
\epsilon_1+\epsilon_1^f+\epsilon_2+\epsilon_2^f+\tau_-+\tau_-^f+\sigma_-+\sigma_-^f+\mu_++\mu_+^f+\mu_-+\mu_-^f,\\
&\Pi_+\times \Xi_-
=
2\epsilon_2+2\epsilon_2^f+2\mu_-+2\mu_-^f,\\
&\Pi_-\times \Pi_-
=
1+f+\epsilon_1+\epsilon_1^f+\tau_++\tau_+^f+\tau_-+\tau_-^f+\sigma_-+\sigma_-^f+\mu_-+\mu_-^f,\\
&\Pi_-\times \Xi_+
=
2\epsilon_2+2\epsilon_2^f+2\mu_++2\mu_+^f,\\
&\Pi_-\times \Xi_-
=
\epsilon_1+\epsilon_1^f+\epsilon_2+\epsilon_2^f+\tau_++\tau_+^f+\sigma_++\sigma_+^f+\mu_++\mu_+^f+\mu_-+\mu_-^f,\\
\eea

\bea
&\Xi_+\times \Xi_+
=
1+f+\epsilon_1+\epsilon_1^f+2\epsilon_2+2\epsilon_2^f+\tau_++\tau_+^f+\tau_-+\tau_-^f+\sigma_++\sigma_+^f+2\sigma_-+2\sigma_-^f+\mu_++\mu_+^f+2\mu_-+2\mu_-^f,\\
&\Xi_+\times \Xi_-
=
2\epsilon_1+2\epsilon_1^f+2\epsilon_2+2\epsilon_2^f+2\mu_++2\mu_+^f+2\mu_-+2\mu_-^f,\\
&\Xi_-\times \Xi_-
=
1+f+\epsilon_1+\epsilon_1^f+2\epsilon_2+2\epsilon_2^f+\tau_++\tau_+^f+\tau_-+\tau_-^f+2\sigma_++2\sigma_+^f+\sigma_-+\sigma_-^f+2\mu_++2\mu_+^f+\mu_-+\mu_-^f.
\eea

\bea
&\Gamma_1\times \bar{\Sigma}_1
=
\Sigma_1+\Sigma_2^f,\\
&\Gamma_1\times \bar{\Sigma}_2
=
\Sigma_1+2\Sigma_2^f+\Sigma_3^f,\\
&\Gamma_1\times \bar{\Sigma}_3
=
\Sigma_2^f+\Sigma_3+2\Sigma_3^f,\\
&\Gamma_2\times \bar{\Sigma}_1
=
\Sigma_1^f+\Sigma_2^f+\Sigma_3,\\
&\Gamma_2\times \bar{\Sigma}_2
=
\Sigma_1+2\Sigma_2+\Sigma_2^f+\Sigma_3+2\Sigma_3^f,\\
&\Gamma_2\times \bar{\Sigma}_3
=
\Sigma_1^f+\Sigma_2+2\Sigma_2^f+3\Sigma_3+2\Sigma_3^f,\\
&\Gamma_3\times \bar{\Sigma}_1
=
\Sigma_1^f+\Sigma_2+\Sigma_3+\Sigma_3^f,\\
&\Gamma_3\times \bar{\Sigma}_2
=
\Sigma_1^f+3\Sigma_2+\Sigma_2^f+3\Sigma_3+2\Sigma_3^f,\\
&\Gamma_3\times \bar{\Sigma}_3
=
\Sigma_1+\Sigma_1^f+3\Sigma_2+2\Sigma_2^f+4\Sigma_3+3\Sigma_3^f,\\
\eea

\bea
&\Omega_+\times \bar{\Sigma}_1
=
\Sigma_2^f+\Sigma_3^f,\\
&\Omega_+\times \bar{\Sigma}_2
=
\Sigma_1+2\Sigma_2^f+\Sigma_3+2\Sigma_3^f,\\
&\Omega_+\times \bar{\Sigma}_3
=
\Sigma_1+\Sigma_2+2\Sigma_2^f+2\Sigma_3+2\Sigma_3^f,\\
&\Omega_-\times \bar{\Sigma}_1
=
\Sigma_2^f+\Sigma_3^f,\\
&\Omega_-\times \bar{\Sigma}_2
=
\Sigma_1+2\Sigma_2^f+\Sigma_3+2\Sigma_3^f,\\
&\Omega_-\times \bar{\Sigma}_3
=
\Sigma_1+\Sigma_2+2\Sigma_2^f+2\Sigma_3+2\Sigma_3^f,\\
\eea

\bea
&\Pi_+\times \bar{\Sigma}_1
=
\Sigma_3+\Sigma_3^f,\\
&\Pi_+\times \bar{\Sigma}_2
=
\Sigma_2+\Sigma_2^f+2\Sigma_3+2\Sigma_3^f,\\
&\Pi_+\times \bar{\Sigma}_3
=
\Sigma_1+\Sigma_1^f+2\Sigma_2+2\Sigma_2^f+2\Sigma_3+2\Sigma_3^f,\\
&\Pi_-\times \bar{\Sigma}_1
=
\Sigma_3+\Sigma_3^f,\\
&\Pi_-\times \bar{\Sigma}_2
=
\Sigma_2+\Sigma_2^f+2\Sigma_3+2\Sigma_3^f,\\
&\Pi_-\times \bar{\Sigma}_3
=
\Sigma_1+\Sigma_1^f+2\Sigma_2+2\Sigma_2^f+2\Sigma_3+2\Sigma_3^f,\\
\eea

\bea
&\Xi_+\times \bar{\Sigma}_1
=
\Sigma_2+\Sigma_2^f+\Sigma_3+\Sigma_3^f,\\
&\Xi_+\times \bar{\Sigma}_2
=
\Sigma_1+\Sigma_1^f+2\Sigma_2+2\Sigma_2^f+3\Sigma_3+3\Sigma_3^f,\\
&\Xi_+\times \bar{\Sigma}_3
=
\Sigma_1+\Sigma_1^f+3\Sigma_2+3\Sigma_2^f+4\Sigma_3+4\Sigma_3^f,\\
&\Xi_-\times \bar{\Sigma}_1
=
\Sigma_2+\Sigma_2^f+\Sigma_3+\Sigma_3^f,\\
&\Xi_-\times \bar{\Sigma}_2
=
\Sigma_1+\Sigma_1^f+2\Sigma_2+2\Sigma_2^f+3\Sigma_3+3\Sigma_3^f,\\
&\Xi_-\times \bar{\Sigma}_3
=
\Sigma_1+\Sigma_1^f+3\Sigma_2+3\Sigma_2^f+4\Sigma_3+4\Sigma_3^f.
\eea

\bea
&\bar{\Sigma}_1\times \bar{\Sigma}_1
=
\Gamma_1^f+\Gamma_2^f+\Gamma_3,\\
&\bar{\Sigma}_1\times \bar{\Sigma}_2
=
\Gamma_1^f+\Gamma_2+\Gamma_3+\Omega_+^f+\Omega_-^f+\Xi_++\Xi_-,\\
&\bar{\Sigma}_1\times \bar{\Sigma}_3
=
\Gamma_2^f+\Gamma_3+\Gamma_3^f+\Omega_+^f+\Omega_-^f+\Pi_++\Pi_-+\Xi_++\Xi_-,\\
&\bar{\Sigma}_2\times \bar{\Sigma}_2
=
2\Gamma_1^f+\Gamma_2+2\Gamma_2^f+3\Gamma_3+\Gamma_3^f+2\Omega_+^f+2\Omega_-^f+\Pi_++\Pi_-+2\Xi_++2\Xi_-,\\
&\bar{\Sigma}_2\times \bar{\Sigma}_3
=
\Gamma_1^f+2\Gamma_2+\Gamma_2^f+3\Gamma_3+2\Gamma_3^f+\Omega_++2\Omega_+^f+\Omega_-+2\Omega_-^f+2\Pi_++2\Pi_-+3\Xi_++3\Xi_-,\\
&\bar{\Sigma}_3\times \bar{\Sigma}_3
=
\Gamma_1+2\Gamma_1^f+2\Gamma_2+3\Gamma_2^f+4\Gamma_3+3\Gamma_3^f+2\Omega_++2\Omega_+^f+2\Omega_-+2\Omega_-^f+2\Pi_++2\Pi_-+4\Xi_++4\Xi_-.
\eea

\end{document}